\documentclass{elsart}

\usepackage[square,comma]{natbib}
\usepackage{graphicx}
\usepackage{pxfonts}

\usepackage{amssymb}

\usepackage{epstopdf}
\begin{document}

\thispagestyle{empty}
\begin{Large}
\textbf{DEUTSCHES ELEKTRONEN-SYNCHROTRON}

\textbf{\large{in der HELMHOLTZ-GEMEINSCHAFT}\\}
\end{Large}

DESY 14-235

December 2014

\begin{eqnarray}
\nonumber &&\cr \nonumber && \cr \nonumber &&\cr
\end{eqnarray}
\begin{eqnarray}
\nonumber
\end{eqnarray}
\begin{center}
\begin{Large}
\textbf{Brightness of synchrotron radiation from wigglers}
\end{Large}
\begin{eqnarray}
\nonumber &&\cr \nonumber && \cr
\end{eqnarray}

\begin{large}
Gianluca Geloni,
\end{large}
\textsl{\\European XFEL GmbH, Hamburg}
\begin{large}

Vitali Kocharyan and Evgeni Saldin
\end{large}
\textsl{\\Deutsches Elektronen-Synchrotron DESY, Hamburg}
\begin{eqnarray}
\nonumber
\end{eqnarray}
\begin{eqnarray}
\nonumber
\end{eqnarray}
ISSN 0418-9833
\begin{eqnarray}
\nonumber
\end{eqnarray}
\begin{large}
\textbf{NOTKESTRASSE 85 - 22607 HAMBURG}
\end{large}
\end{center}
\clearpage
\newpage

\begin{frontmatter}


\title{Brightness of synchrotron radiation from wigglers}


\author[XFEL]{Gianluca Geloni}
\author[DESY]{Vitali Kocharyan}
\author[DESY]{Evgeni Saldin}

\address[XFEL]{European XFEL GmbH, Hamburg, Germany}
\address[DESY]{Deutsches Elektronen-Synchrotron (DESY), Hamburg,
Germany}

\begin{abstract}
According to literature, while calculating  the brightness of synchrotron radiation from wigglers, one needs to account for the so called `depth-of-field' effects. In fact, the particle beam cross section varies along the wiggler. It is usually stated that the effective photon source size increases accordingly, while the brightness is reduced. Here we claim that this is a misconception originating from an analysis of the wiggler source based on geometrical arguments, regarded as almost self-evident. According to electrodynamics, depth-of-field effects do not exist: we demonstrate this statement both theoretically and  numerically, using a well-known first-principle computer code. This fact shows that under the usually accepted approximations, the description of the wiggler brightness turns out to be inconsistent even qualitatively. Therefore, there is a need for a well-defined procedure for computing the brightness from a wiggler source. We accomplish this task based on the use of a Wigner function formalism. In the geometrical optics limit computations can be performed analytically. Within this limit, we restrict ourselves to the case of the beam size-dominated regime, which is typical for synchrotron radiation facilities in the X-ray wavelength range. We give a direct demonstration of the fact that the apparent horizontal source size is broadened in proportion to the beamline opening angle and to the length of the wiggler. While this effect is well-understood, a direct proof appears not to have been given elsewhere. We consider the problem of the calculation of the wiggler source size by means of numerical simulations alone, which play the same role of an experiment. We report a significant numerical disagreement between exact calculations and approximations currently used in literature.
\end{abstract}


\end{frontmatter}


\clearpage

\section{\label{sec:intro} Introduction}

The magnetic system of wigglers is identical to that of undulators. Both devices generate a periodic magnetic field to enhance the radiation intensity. However, while a typical undulator is characterized by a moderate deflection parameter $K \lesssim 3$, a wiggler is usually endowed with a much larger $K$ value in the range $10-30$. Such difference in $K$ values causes a significant difference in the spectrum of the radiation produced by these two devices.  In fact, the number of harmonics in the spectrum increases significantly as the $K$ values increases. As a result, the spectrum of the radiation from undulators usually includes harmonics from the 1st to 7th, while wigglers operate up to a much higher harmonic number $n$, with $n > 100$.  Since the phase errors in the radiation field are proportional to the harmonic number, they are much more severe in the wiggler case than in the undulator case. Because of smaller phase errors, undulators achieve their high angular flux density and quasi-monochromatic spectrum due to interference of the radiation emitted by the different magnetic poles. In the case of wigglers instead, due to larger phase errors and electron beam energy spread, the interference pattern is usually smoothed out.  Thus, the properties of a wiggler source are usually calculated as that for bending magnets, and radiation is incoherently summed up over all the periods. In this incoherent model, the flux distribution is given by the appropriate bending magnet expression multiplied by $2N$, where $N$ is the number of magnetic periods in the device \cite{HUL1,HUL2, WALK}.

In the past, most experimental applications of wiggler radiation benefited from high flux, and it was important to quantify only the angular spectral flux density radiated by the source. However, at present, applications such as macromolecular crystallography and micro-diffraction makes it important to quantify not only the radiation flux, but also the brightness of a
wiggler source \cite{SHAFT}.

The physical meaning of brightness can be best understood by considering the imaging of the source on a given experimental sample. The brightness is a figure of merit that quantifies how well a SR beam can be focused. However, in general, the maximum photon flux density onto the image plane is altered by optical elements along the setup. The brightness can be interpreted as the theoretically maximal concentration of the SR photon flux on the image-receiving surface where, usually, the sample is placed.

According to literature, while calculating  the brightness of a wiggler source one needs to take into account depth-of-field effects, that are the contribution to the apparent source size from different poles. From electron beam dynamics considerations it is known that when focusing elements are absent the electron beam cross sections vary along the wiggler according to

\begin{eqnarray}
\sigma_{x,y}^2(z) = \sigma_{x,y}^2(0) + z^2 \sigma_{x',y'}^2 ~,
\label{xsecxy}
\end{eqnarray}
where $-L/2 < z < L/2$ is the distance from the waist,  $L$ is the wiggler length, $\sigma_{x, y}$ are the electron beam rms sizes and $\sigma_{x',y'}$ the electron beam rms divergences at the waist. Authors of papers \cite{HUL1,HUL2} and review \cite{WALK} regarded the changing  of the rms electron beam sizes along the wiggler as a clear evidence of the fact that the photon source size changes  along the wiggler as well. Consequently,  since the average particle beam size at a given position down the wiggler is larger than that at the waist, one observes an increase of the effective source size and a reduction of the wiggler brightness. However, we hold this notion for a misconception: we will show that, according to the laws of electrodynamics of ultra-relativistic charged particles, the photon source size is not widened at all and the expression for the brightness only includes the electron beam size at the waist. This fact demonstrates the need for a novel formulation of the theory of  brightness for wiggler sources. In this article we consider such a theory based on a Wigner distribution formalism. In Section \ref{sec:wigwig} we develop the theory. In section \ref{sec:overview}  we give a summary of results in literature, and we discuss an overview of novel findings. Numerical examples will be discussed in the following Section in order to demonstrate geometrical properties of the wiggler source.

Throughout this work, the source is considered placed in the middle of the wiggler, and is obviously virtual in nature. However, this is no mathematical abstraction. In fact, if we take a single focusing mirror to form a 1:1 image and set the object plane in the middle of the wiggler, the image obtained is a visualization of the virtual source. Simple cases pertaining the geometrical optics limit can be calculated analytically. In this paper we will illustrate, as an example, the case of the beam size-dominated regime. More in general, one can note that first-principle computer codes (see e.g. SRW \cite{CHUB} and SPECTRA \cite{SPECTRA}) have been used quite successfully to model advanced SR source and beamlines without specific analytical simplifications. These codes can be used to treat the case of 1:1 imaging of a wiggler source as well. It is possible to calculate the wiggler source by means of numerical simulations alone, which play the same role of an experiment. One can check that results of simulations confirm our prediction that the image is insensitive to the electron beam divergence. In contrast with the prediction \cite{HUL1,HUL2, WALK} of significant source size widening we find, with graphical accuracy, that the distribution at image plane remains unvaried while increasing the electron beam divergence.

At variance, as is well understood, the apparent horizontal source size is broadened proportionally to the beamline opening angle and to the wiggler length \cite{WALK}, but a direct demonstration of this effect appears not to have been provided \cite{BERM}. In this article we consider the problem of calculating the wiggler source size as a function of the beamline angular acceptance by means of numerical simulations. Based on the use of the code SRW we report a significant numerical disagreement between our exact calculations and approximations found in literature,  which are based on the use geometrical optics.


\section{\label{sec:wigwig} Wigner distribution and wiggler source}

The brightness is an appropriate figure of merit for estimating the photon flux density on a given sample. It was originally defined with the help of traditional radiometry. Traditional radiometry follows a geometrical-optics approach, and provides a natural description of the properties of light from incoherent sources such as second generation synchrotron radiation (SR) sources. The basic quantity in this discipline is in fact the radiance, that is the photon flux density in phase space. For incoherent sources the brightness is nothing more than the maximum value of the radiance. An extension of the concept of brightness beyond the realm of geometrical optics, for example to third generation SR sources, requires a redefinition in terms of electromagnetic fields and their statistical properties, based on classical relativistic electrodynamics and statistical optics. Such generalization was first proposed by Kim \cite{KIM1,KIM3} and relies on the Wigner distribution (WD) \cite{WIGN} of the SR electric fields. In literature, the brightness of a SR source is sometimes defined as the Wigner distribution itself, i.e as a quasi-probability distribution in phase-space \cite{KIM1,KIM3,CIOC,ELLE}. As such, for a SR source of arbitrary state of transverse coherence, it is not guaranteed to be positive everywhere. Moreover it is convenient to introduce a figure of merit which always gives back a single, positive number and can serve, at the same time, as measure of the WD. Finally a particular correspondence principle should be satisfied, is based on the assumption that the formalism involved in the calculation of the brightness must include radiometry as a limiting case.

We consider the maximum of the Wigner distribution (WD) of synchrotron radiation (SR) fields as a definition of SR source brightness \cite{OURJSR}. Such figure of merit constitutes a generalization of Kim's choice of the on-axis peak value of the WD as a figure of merit for the undulator case \cite{KIM1,KIM3}. The brightness defined in this way is always positive and, in the geometrical optics limit, it can be interpreted as the maximum of the photon flux density in phase space, that is the maximum of the radiance. Then, the correspondence principle mentioned above consists in using the classical definition of radiance to obtain a correct proportionality factor in the definition of brightness \cite{OURJSR}.

\subsection{Radiation emitted by a single electron}

Let us first describe a general method to calculate the WD of SR emitted by a single electron. We will be interested in the case of an ultra-relativistic electron going through a certain magnetic system. We will discuss of a wiggler in order to illustrate our reasoning, but the consideration in this section is fully general, and applies to any other magnetic system like undulators and bending magnets as well. SR theory is naturally developed in the space-frequency domain, as one is usually interested in radiation properties at a given position in space and at a certain frequency. In this article we define the relation between temporal and frequency domain via the following definition of Fourier transform pair:

\begin{eqnarray}
&&\bar{f}(\omega) = \int_{-\infty}^{\infty} dt~ f(t) \exp(i \omega t ) \leftrightarrow
f(t) = \frac{1}{2\pi}\int_{-\infty}^{\infty} d\omega \bar{f}(\omega) \exp(-i \omega t) ~.
\label{ftdef2}
\end{eqnarray}
%

When one needs to specify the Wigner distribution at any position down the beamline, one needs to calculate the field at any position down the beamline as well. In order to do so, we first calculate the field from a single electron moving along an arbitrary trajectory in the far zone, and then we solve the propagation problem in paraxial approximation. This last step allows us to calculate the field at any position by backward-propagation in free-space with the help of the paraxial Green's function, that is the Fresnel propagator.

We call $z$ the observation distance along the optical axis of the system, while $\vec{r}$ fixes the transverse position of the observer. Suppose we are interested in the radiation generated by an electron and observed far away from it. In this case it is possible to find a relatively simple expression for the electric field \cite{JACK}. We indicate the electron velocity in units of $c$ with $\vec{\beta}$, the Lorentz factor (that will be considered fixed throughout this paper) with $\gamma$, the electron trajectory in three dimensions with $\vec{R}(t)$ and the observation position with $\vec{R}_0 = (z_0,r_0)$. Finally, we introduce the unit vector

\begin{equation}
\vec{n} =
\frac{\vec{R}_0-\vec{R}(t)}{|\vec{R}_0-\vec{R}(t)|}
\label{enne}
\end{equation}
%

pointing from the retarded position of the electron to the observer. In the far zone, by definition, the unit vector $\vec{n}$ is nearly constant in time. If the position of the observer is far away enough from the charge, one can make the expansion

\begin{eqnarray}
\left| \vec{R}_0-\vec{R}(t) \right|= R_0 - \vec{n} \cdot \vec{R}(t)~.
\label{next}
\end{eqnarray}
We then obtain the following approximate expression for the the radiation field  in the space-frequency domain\footnote{For a better understanding of the physics involved one can refer to e.g. the textbook \cite{JACK}. A different constant of proportionality in Eq. (\ref{revwied}) is to be ascribed to the use of different units and definition of the Fourier transform.}:

\begin{eqnarray}
\vec{\bar{E}}(\vec{R}_0,\omega) &=& -{i\omega e\over{c
R_0}}\exp\left[\frac{i \omega}{c}\vec{n}\cdot\vec{R}_0\right]
\int_{-\infty}^{\infty}
dt~{\vec{n}\times\left[\vec{n}\times{\vec{\beta}(t)}\right]}\exp
\left[i\omega\left(t-\frac{\vec{n}\cdot
\vec{R}(t)}{c}\right)\right] \cr &&  \label{revwied}
\end{eqnarray}
where $\omega$ is the frequency, $(-e)$ is the negative electron charge and we make use of Gaussian units. Using the complex notation, in this and in the following sections we assume, in agreement with Eq. (\ref{ftdef2}), that the temporal dependence of fields with a certain frequency is of the form:

\begin{eqnarray}
\vec{E} \sim \vec{\bar{E}}(z,\vec{r},\omega) \exp(-i \omega t)~.
\label{eoft}
\end{eqnarray}
With this choice for the temporal dependence we can describe a plane wave traveling along the positive  $z$-axis with

\begin{eqnarray}
\vec{E} = \vec{E}_0 \exp\left(\frac{i\omega}{c}z -i \omega t\right)~.
\label{eoftrav}
\end{eqnarray}
In the following we will always assume that the ultra-relativistic approximation is satisfied, which is the case for SR setups. As a consequence, the paraxial approximation applies too. The paraxial approximation implies a slowly varying envelope of the field with respect to the wavelength. It is therefore convenient to introduce the slowly varying envelope of the transverse field components as

\begin{equation}
\vec{\widetilde{E}}(z,\vec{r},\omega) = \vec{\bar{E}}(z,\vec{r},\omega) \exp{\left(-i\omega z/c\right)}~. \label{vtilde}
\end{equation}
Introducing angles $\theta_x = x_0/z_0$ and $\theta_y = y_0/z_0$, the transverse components of the envelope of the field in Eq. (\ref{revwied}) in the far zone and in paraxial approximation can be written as

\begin{eqnarray}
\vec{\widetilde{{E}}}(z_0, \vec{r}_0,\omega) &=& -{i
\omega e\over{c^2}z_0} \int_{-\infty}^{\infty} dz' {\exp{\left[i
\Phi_T\right]}}  \left[\left({v_x(z')\over{c}}
-\theta_x\right){\vec{e_x}}
+\left({v_y(z')\over{c}}-\theta_y\right){\vec{e_y}}\right] \label{generalfin}
\end{eqnarray}
where the total phase $\Phi_T$ is

\begin{eqnarray}
&&\Phi_T = \omega \left[{s(z')\over{v}}-{z'\over{c}}\right] \cr &&
+ \frac{\omega}{2c}\left[z_0 (\theta_x^2+\theta_y^2) - 2 \theta_x x(z') - 2 \theta_y y(z') + z'(\theta_x^2+\theta_y^2)\right]~
. \label{totph}
\end{eqnarray}
Here $v_x(z')$ and $v_y(z')$ are the horizontal and the vertical components of the transverse velocity of the electron,  $x(z')$ and $y(z')$ specify the transverse position of the electron as a function of the longitudinal position, $\vec{e}_x$ and $\vec{e}_y$ are unit vectors along the transverse coordinate axis. Finally, $s(z')$ is the longitudinal coordinate along the trajectory. The electron is moving with velocity $\vec{v}$, whose magnitude is constant and equal to $v = ds/dt$.

Eq. (\ref{generalfin}) can be used to characterize the far field from an electron moving on any trajectory.  When the single-electron fields are specified at a certain position $z_1$, the fields at any other position $z_2$ can be found by propagating forward or backward in free-space according to the paraxial law

\begin{eqnarray}
&&\widetilde{E} \left(z_2,\vec{r}_{2}, \omega\right) = \frac{i \omega}{2
\pi c( {z}_2- {z}_1)} \int d \vec{ {r}_1}~\widetilde{E} \left(z_1,\vec{r}_{1}, \omega\right) \exp{\left[\frac{i \omega
\left|{\vec{ {r}}}_{2}-\vec{ {r}_{1}}\right|^2}{2 c (
{z}_2- {z}_1)}\right]}~. \cr && \label{fieldpropback}
\end{eqnarray}
In particular, one may decide to backpropagate the field even at positions well inside the magnetic structure under study. In this case, the field distribution is obviously virtual in nature, because it is not actually there, but it fully characterizes the radiation field from a single electron. Within the paraxial approximation, single-electron fields are fully characterized when they are known on a transverse plane at one arbitrary position $z$. Because of this, all positions $z$ are actually equivalent.  Without loss of generality one can set $z_s =0$ for simplicity and call this the source position. Then, the relation between the field from a single electron at the source $\widetilde{E} \left(0,\vec{r}, \omega\right)$  and the field in the far zone, $\widetilde{E} \left(z_0,\vec{\theta}, \omega\right) $, follows once more from Eq. (\ref{fieldpropback}):

\begin{eqnarray}
&&\widetilde{E} \left(0,\vec{r}, \omega\right) = \frac{i z_0 \omega}{2\pi c} \int d \vec{\theta}~ \widetilde{E} \left(z_0,\vec{\theta}, \omega\right) \exp\left(-\frac{i\theta^2 z_0 \omega}{2c}\right)\exp \left(\frac{i \omega \vec{r}\cdot\vec{\theta}}{c}\right)~\cr &&
\label{farzone}
\end{eqnarray}
\begin{eqnarray}
&&\widetilde{E} \left(z_0,\vec{\theta}, \omega\right) = \frac{i \omega}{2\pi c z_0} \exp\left(\frac{i\theta^2 z_0 \omega}{2c}\right) \int d \vec{r} ~\widetilde{E} \left(0,\vec{r}, \omega\right)  \exp \left(-\frac{i \omega\vec{r}\cdot\vec{\theta}}{c}\right)\cr &&
\label{farzonef}
\end{eqnarray}
We assume that a plane wave traveling along the positive $z$-axis can be expressed as in Eq. (\ref{eoftrav}). Then, the negative sign in the exponential factor $\exp(-i \omega z/c)$ in Eq. (\ref{vtilde}) determines the sign of the exponential in Eq. (\ref{fieldpropback}) and consequently the sign of the exponential that appears in the integrand in Eq. (\ref{farzonef}), which is the solution of the propagation problem in the far zone.

Let us now discuss the case of wiggler radiation from a single electron with an arbitrary angular deflection $\vec{\eta}$ and an arbitrary offset $\vec{l}$ with respect to a reference orbit defined as the path through the origin of the coordinate system, that is $x(0) = y(0) = 0$.

If the magnetic field in the setup does not depend on the transverse coordinates, i.e. $B = B(z)$, an initial offset $x(0) = l_x$, $y(0) = l_y$ shifts the trajectory of an electron of $\vec{l}$. Similarly, an angular deflection $\vec{\eta} = (\eta_x, \eta_y)$ at $z=0$ tilts the trajectory without modifying it. Cases when the magnetic field of SR sources include focusing elements (or the natural focusing of insertion devices) are out of the scope of this paper. Assuming further that $\eta_x \ll 1$ and $\eta_y \ll 1$, which is typically justified for ultrarelativistic electron beams, one obtains the following approximation for the electron trajectory:

\begin{eqnarray}
&& \vec{r}(z) = \vec{r}_r(z) + \vec{\eta} z + \vec{l} ~, \cr &&
\vec{v}(z) = \vec{v}_r(z) + v \vec{\eta}~,
\label{shiftilt}
\end{eqnarray}
where the subscript `r' refers to the reference trajectory. The pair $(\vec{r}(z),z)$  gives a parametrical description of the trajectory of a single electron with offset $\vec{l}$ and deflection $\vec{\eta}$. The curvilinear abscissa on the trajectory can then be written as

\begin{eqnarray}
&& s(z) = \int_0^z dz' \left[1+\left(\frac{dx}{dz'} \right)^2+\left(\frac{dy}{dz'} \right)^2\right]^{1/2} \cr &&
\simeq \int_0^z dz' \left[1 + \frac{1}{2} \left(\frac{dx_r}{dz'} \right)^2+\frac{1}{2}\left(\frac{dy_r}{dz'}\right)^2 + \frac{1}{2} \left(\eta_x^2 + \eta_y^2\right) + \eta_x \frac{d x_r}{dz'} + \eta_y \frac{d y_r}{dz'}\right] \cr &&= s_r(z) + \frac{\eta^2 z}{2} + \vec{r}_r(z) \cdot \vec{\eta}~,
\label{curvabs}
\end{eqnarray}
where we expanded the square root around unity in the first passage, we made use of Eq. (\ref{shiftilt}), and of the fact that the curvilinear abscissa along the reference trajectory is $s_r(z) \simeq z + \int_0^z |d \vec{r}_r/dz'|^2/2$.

We now substitute Eq. (\ref{shiftilt}) and Eq. (\ref{curvabs}) into Eq. (\ref{generalfin}) to obtain:

\begin{eqnarray}
&& \vec{\widetilde{{E}}}(z_0, \vec{r}_0,\omega) = -{i
\omega e\over{c^2}z_0} \int_{-\infty}^{\infty} dz' {\exp{\left[i
\Phi_T\right]}} \cr && \times  \left[\left({v_x(z')\over{c}}
-(\theta_x-\eta_x)\right){\vec{e_x}}
+\left({v_y(z')\over{c}}-(\theta_y-\eta_y)\right){\vec{e_y}}\right]
~,\cr && \label{generalfin2}
\end{eqnarray}
where the total phase $\Phi_T$ is

\begin{eqnarray}
&&\Phi_T = \omega \left[\frac{s_r(z')}{v} +\frac{\eta^2 z'}{2 v} + \frac{1}{v}\vec{r}_r(z') \cdot \vec{\eta} -{z'\over{c}}\right] \cr &&
+ \frac{\omega}{2c}\left[z_0 (\theta_x^2+\theta_y^2) - 2 \theta_x x_r(z')  - 2 \theta_x \eta_x z' - 2 \theta_x l_x \right. \cr &&  \left. - 2 \theta_y y(z') - 2 \theta_y \eta_y z' - 2 \theta_y l_y  + z'(\theta_x^2+\theta_y^2)\right]~,
 \label{totph2}
\end{eqnarray}
which can be rearranged as

\begin{eqnarray}
&&\Phi_T \simeq
\omega \left[{s_r(z')\over{v}}-{z'\over{c}}\right]  - \frac{\omega}{c} (\theta_x l_x + \theta_y l_y) \cr &&
+ \frac{\omega}{2c}\left[z_0 (\theta_x^2+\theta_y^2) - 2 (\theta_x-\eta_x) x_r(z') \right. \cr && \left. - 2 (\theta_y-\eta_y) y_r(z') + z'\left((\theta_x-\eta_x)^2+(\theta_y-\eta_y)^2\right)\right] ~.\cr &&
\label{totph3}
\end{eqnarray}

The expression for the field at virtual source should be modified accordingly. Namely, one should plug Eq. (\ref{generalfin2}) and Eq. (\ref{totph3}) into Eq. (\ref{farzone}), which gives

\begin{eqnarray}
\widetilde{E}\left(\vec{l},\vec{\eta},0,\vec{r},\omega\right) = \widetilde{E}_0\left(\vec{r}-\vec{l}\right)\exp\left[i \omega \vec{\eta} \cdot \left(\vec{r}-\vec{l}\right)/c \right]
\label{tiltshift}
\end{eqnarray}
where we set $\widetilde{E}_0\left(\vec{r}\right)\equiv \widetilde{E}\left(0,0,0,\vec{r},\omega\right)$. The presence of an electron offset $\vec{l}$ shifts the single-electron field source, while a deflection $\eta$ tilts the source.

\subsection{Electron bunch effects}

Since SR is a random process, the description of properties of the source should be treated in terms of probabilistic statements. In fact, the shot noise in the electron beam causes fluctuations of the electron beam current density. These fluctuations are random both in space and time. Statistical optics \cite{GOOD} provides the most convenient tools to deal with fluctuating electromagnetic fields.  From the viewpoint of statistical optics, synchrotron radiation is a Gaussian random process\footnote{This fact is commonly known and accepted in the synchrotron radiation community. However, we have been unable to find in literature an explicit proof except in reference \cite{OURTR}, which is the only paper, to the authors' knowledge, dealing with this issue.}. An important consequence of this fact is that higher-order correlation functions can be expressed in terms of the second order correlation with the help of the moment theorem \cite{GOOD}. As a result, the knowledge of the second order correlation function  is we needs in order to completely characterize the signal from a statistical viewpoint. Due to a limited temporal resolution of detectors in SR experiments, the analysis in space-frequency domain is much more natural than that in the space-time domain. As a consequence of this choice, we study the spatial correlation between Fourier transform of the electric field at the fixed frequency that is the cross-spectral density

\begin{equation}
G\left(z,\vec{r}+\frac{\Delta \vec{r}}{2} , \vec{r}-\frac{\Delta \vec{r}}{2} , \omega\right) \equiv \left\langle
\widetilde{E} \left(\vec{\eta},\vec{l},z,\vec{r}+\frac{\Delta \vec{r}}{2} , \omega\right)
\widetilde{E}^*\left(\vec{\eta},\vec{l},z,\vec{r}-\frac{\Delta \vec{r}}{2} , \omega\right)
\right\rangle~,\label{coore}
\end{equation}
where the brackets $\langle ... \rangle$ indicate ensemble average over electron bunches.

Let us consider a certain phase space distribution for an electron beam with a given transverse phase space distribution $f_\bot(\vec{l},\vec{\eta})$, which is a function of offset $\vec{l}$ and deflection $\vec{\eta}$. At the source position one can write

\begin{eqnarray}
G\left(\vec{r}, \Delta \vec{r}\right) = \int d\vec{l}~d\vec{\eta}~f_\bot\left(\vec{l},\vec{\eta}\right)  \widetilde{E}\left(\vec{l},\vec{\eta},\vec{r}+ \frac{\Delta \vec{r}}{2}\right) \widetilde{E}^*\left(\vec{l},\vec{\eta},\vec{r} - \frac{\Delta \vec{r}}{2}\right) ~,
\label{GGG2}
\end{eqnarray}
where, for notation simplicity, we did not indicate the dependence of the single particle fields on $\omega$. There are practical situations when offset and deflection of an electron lead to the same offset and deflection of the radiation beam from that electron. This is the case for magnetic setups as undulators, bending magnets and wigglers without focusing elements. In such situations from Eq. (\ref{tiltshift}) one obtains

\begin{eqnarray}
G\left(\vec{r},\Delta \vec{r}\right) = \int d \vec{l} ~ G_0\left(\vec{r}-\vec{l},\Delta \vec{r}\right)  \int d\vec{\eta} f_\bot\left(\vec{l},\vec{\eta}\right) \exp\left(i \omega \vec{\eta}\cdot \Delta \vec{r}/c\right)
\label{GG0}
\end{eqnarray}
where $G_0(\vec{r}, \Delta \vec{r}) \equiv \widetilde{E}_0(\vec{r}+\Delta \vec{r}/2)\widetilde{E}_0^*(\vec{r}-\Delta \vec{r}/2)$.

Aside for the normalization constant $\mathcal{A}$, which will be defined later in Eq. (\ref{norm}), the inverse Fourier transform of the cross-spectral density with respect to $\Delta x$ and $\Delta y$ can be written as:

\begin{eqnarray}
W(\vec{r}, \vec{\theta}) = \mathcal{A} \int d\Delta\vec{r} ~G(\vec{r}, \Delta \vec{r})  \exp\left(-i \omega \vec{\theta} \cdot \Delta \vec{r}/c \right)~.
\label{Wig1}
\end{eqnarray}
This is the expression for the Wigner distribution in terms of the cross-spectral density. We regard it as a distribution function defined over the four dimensions $(\vec{r},\vec{\theta})$ and parameterized by $z$. It can be shown that $W$ always assumes real values, although it is not always a positive function. Yet, the integral over $\vec{r}$ and $\vec{\theta}$ can be shown to be positive, and therefore the maximum of the Wigner distribution is also bound to be positive, so that we can take this value as a natural definition of the brightness of SR sources \cite{OURJSR}.

The basic properties of the Wigner distribution include statements about its 2D projections. In particular, if we make use of Eq. (\ref{Wig1}) we obtain the following expression for the $(x,y)$ projection:

\begin{eqnarray}
\int d\vec{\theta} ~W(\vec{r}, \vec{\theta}) = (2\pi)^2 \frac{c^2}{\omega^2}\mathcal{A} \left \langle \left| \bar{E}(\vec{r})\right|^2 \right \rangle~.
\label{partialint}
\end{eqnarray}
In the geometrical optics limit, $W$ can be interpreted as the photon distribution in phase space. Then, for consistency with this limit, one should require that integrating the Wigner distribution function over the solid angle $d\Omega = d \vec{\theta}$ yields the  photon spectral and spatial flux density: 

\begin{eqnarray}
\int d\vec{\theta}~ W(\vec{r}, \vec{\theta}) = \frac{d \dot{N}_{ph}}{dS (d\omega/\omega)} = \frac{I}{e \hbar}  \frac{c}{4 \pi^2} \left\langle\left|\bar{E}(\vec{r})\right|^2 \right \rangle ~,
\label{intW}
\end{eqnarray}
where $I$ is  electron beam current, $e$ is charge of the electron taken without sign, $c$ is the speed of light in vacuum and $\hbar = h/(2\pi)$ is the reduced Planck constant.  Here we have used Parseval theorem, and included an additional factor two on the right-hand side of Eq. (\ref{intW}),  indicating that we use positive frequencies only. Comparison of the requirement in Eq. (\ref{intW}) with the mathematical property in Eq. (\ref{partialint}) fixes univocally the normalization constant $\mathcal{A}$ as

\begin{eqnarray}
\mathcal{A} = \frac{c}{(2 \pi)^4} \frac{I}{e \hbar} \left(\frac{\omega}{c}\right)^2 ~.
\label{norm}
\end{eqnarray}
Note that $\mathcal{A}$ depends on the units chosen (in this case Gaussian units) and on our definition of Fourier transformation (Eq. (\ref{ftdef2})). The Wigner distribution $W$ is also univocally defined as

\begin{eqnarray}
W\left(\vec{r}, \vec{\theta}\right) = \frac{c}{(2 \pi)^4} \frac{I}{e \hbar} \left(\frac{\omega}{c}\right)^2 \int d\Delta\vec{r} ~G(\vec{r}, \Delta \vec{r})  \exp\left(-i \omega \vec{\theta} \cdot \Delta \vec{r}/c \right)~.
\label{Wig13}
\end{eqnarray}
In the presence of electron bunch effects a very useful  addition theorem  can be obtained \cite{KIM1}. This theorem is commonly known and accepted in the synchrotron radiation community and, in particular, it is applied by code writers for numerical calculations of the Wigner distribution \cite{BAZA,TANA}. However,  we have been unable to find in the literature an explicit proof of its  application to the present case. Such a proof can be given in the following way. If the magnetic field of the SR source does not depend on the transverse position, then we can express the cross spectral density as Eq. (\ref{GG0}). Substituting Eq. (\ref{GG0}) into Eq. (\ref{Wig13}) one obtains

\begin{eqnarray}
W\left(\vec{r},\vec{\theta}\right) =  \int d \vec{l} d\vec{\eta} ~W_0\left(\vec{r}-\vec{l}, \vec{\theta} - \vec{\eta} \right)   f_\bot\left(\vec{l},\vec{\eta}\right)
\label{Wcorr}
\end{eqnarray}
with $W_0$ defined as the Wigner distribution associated to $G_0$. This can be summarized by saying that the electron offset and deflection correspond to an offset in position and angle of the corresponding Wigner distribution $W_0$, and that the overall Wigner distribution $W$ can be found by addition over single-electron contributions.

We now turn to the main topic of our study, namely the analysis of the brightness of a wiggler source. It is our purpose here to demonstrate how a straightforward application of Eq. (\ref{Wcorr}) yields analytical expressions for the brightness in the geometrical optics limits. The Wigner distribution $W\left(\vec{r},\vec{\theta}\right)$ for an electron beam with finite emittance can be presented as a convolution product between the electron phase space distribution $f_\bot\left(\vec{l},\vec{\eta}\right)$ and the Wigner distribution for a filament beam $W_0\left(\vec{r},\vec{\theta}\right)$ according to Eq. (\ref{Wcorr}). It should be noted that, as remarked before, Eq. (\ref{Wcorr}) can only be used in the case when focusing elements are excluded from consideration.

We assume that the motion of electrons in the horizontal and vertical directions are completely uncoupled. Additionally, we assume a Gaussian distribution for the electron beam phase-space. These two assumptions are practically realized, with good accuracy, in storage rings. For simplicity we also assume that the minimal values of the beta-functions in horizontal and vertical directions are located at the middle of the wiggler, at $z = 0$.  Then, at that position, the transverse phase-space distribution can be expressed as

\begin{eqnarray}
{f}_\bot = f_{\vec{l}}\left(\vec{l}\right)f_{\vec{\eta}}\left(\vec{\eta}\right)=f_{\eta_x}(\eta_x) f_{\eta_y}(\eta_y)
f_{l_x}(l_x) f_{l_y}(l_y)
\label{fphase}
\end{eqnarray}
with

\begin{eqnarray}
&& f_{\eta_x}(\eta_x) = \frac{1}{\sqrt{2\pi}\sigma_{x'}} \exp\left(-\frac{\eta_x^2} {2\sigma_{x'}^2}\right) ~,~~~~
f_{\eta_y}(\eta_y) = \frac{1}{\sqrt{2\pi}\sigma_{y'}} \exp\left(-\frac{\eta_y^2} {2\sigma_{y'}^2}\right) ~,\cr &&
f_{l_x}(l_x) =\frac{1}{\sqrt{2\pi} \sigma_{x} } \exp\left(-\frac{{l}_x^2}{2 \sigma_{x}^2}\right)~,~~~~~
f_{l_y}(l_y) =\frac{1}{\sqrt{2\pi} \sigma_{y} } \exp\left(-\frac{{l}_y^2}{2 \sigma_{y}^2}\right)~.\cr && \label{distr}
\end{eqnarray}

We begin by writing the expression for the cross-spectral density at the virtual source:

\begin{eqnarray}
&& G\left(\vec{{r}}, \Delta \vec{{r}}\right) \cr && = \int d \vec{{\eta}} \exp\left(i  \vec{{\eta}}\cdot \Delta \vec{{r}}\right) f_{\vec{{\eta}}}\left(\vec{{\eta}}\right) \int d \vec{{l}} f_{\vec{{l}}}\left(\vec{{l}}\right)
\widetilde{E}_0\left(\vec{{r}} + \frac{\Delta \vec{{r}}}{2} -\vec{{l}}\right) \widetilde{E}_0^*\left(\vec{{r}} -\frac{\Delta \vec{{r}}}{2} - \vec{{l}}\right) ~,\cr &&
\label{G0norm}
\end{eqnarray}
where the field is defined by Eq. (\ref{tiltshift}). One sees that the cross-spectral density is the product of two separate factors. The first is the Fourier transform of the distribution of the electrons angular divergence. The second is the convolution of the transverse electron beam distribution with the four-dimensional function $\widetilde{E}\left(\vec{{r}} + {\Delta \vec{{r}}}/{2} \right) \widetilde{E}^*\left(\vec{{r}} -{\Delta \vec{{r}}}/{2} \right)$. In fact, after the change of variables $\vec{\phi} = \vec{r} - \vec{l}$ we have

\begin{eqnarray}
G\left(\vec{{r}},\Delta \vec{r}\right) &=&
\frac{1}{2 \pi \sigma_x\sigma_y} \exp \left[-\frac{\omega^2 \Delta {x}^2
\sigma_{x'}^2}{2 c^2}\right] \exp \left[-\frac{\omega^2 \Delta {y}^2 \sigma_{y'}^2}{2 c^2}\right]\cr &&
\times \int_{-\infty}^{\infty} d \phi_x \int_{-\infty}^{\infty} d
\phi_y \exp\left[-\frac{\left(\phi_x-{x}\right)^2}{2
\sigma_x^2}\right] \exp\left[-\frac{\left(\phi_y-{y}\right)^2}{2
\sigma_y^2}\right] \cr && \times
\widetilde{E}_0\left({\phi_{x}}+\frac{\Delta {x}}{2},{\phi_{y}}+\frac{\Delta {y}}{2}
\right) \widetilde{E}_0^*\left({\phi_{x}}-\frac{\Delta {x}}{2}
,{\phi_{y}}-\frac{\Delta {y}}{2}
\right)~.\label{Gnor3}
\end{eqnarray}
This expression for the cross-spectral density is obtained by evaluating integrals over deflection angles using only fact that the single electron field distribution at the virtual source satisfies Eq. (\ref{tiltshift}). This means that Eq. (\ref{Gnor3}) must be true for any magnetic setup (e.g. bending magnet, undulator, wiggler) without focusing elements.

It is instructive to examine this expression in the geometrical optics asymptotes. Let us start with the beam size-dominated regime, which is typical for SR facilities in the X-ray wavelength range.  From a mathematical viewpoint we can discuss the asymptotic\footnote{This assumption and others like the restriction to an electron beam without energy spread, the fact that the electron beam waist is located at the virtual source position and the fact that the electron phase space can be factorized in product of factors separately including size and divergence are often not met in practice. However, they present several advantages. First, they allow for a more transparent analytical treatment. Moreover, they permit a direct comparison of results with books and articles, where similar assumptions are made. Finally, these results can still be used to compare the performance of facilities, which should be done by producing a single number, a figure of merit, at the same conventionally chosen working conditions. One can then further proceed with a generalization to fully realistic situations applying numerical techniques as done, for example, in \cite{TANA}, and available in the code SPECTRA \cite{SPECTRA}.} $\sigma_{x,y} \longrightarrow \infty$ and $\sigma_{x',y'} \longrightarrow 0$. We thus obtain

\begin{eqnarray}
&&G = \frac{1}{2\pi \sigma_x \sigma_y} \exp\left( - \frac{x^2}{2\sigma_x^2}\right) \exp\left( - \frac{y^2}{2\sigma_y^2}\right) \cr && \times \int_{-\infty}^{\infty} d x' \int_{-\infty}^{\infty} dy' \widetilde{E}_0\left( x'+\frac{\Delta x}{2}, y'+\frac{\Delta y}{2} \right) \widetilde{E}_0^*\left( x'-\frac{\Delta x}{2}, y'-\frac{\Delta y}{2}\right)~. \cr &&
\label{Gthird}
\end{eqnarray}
The Wigner distribution in Eq. (\ref{Wig13}) can therefore be written as

\begin{eqnarray}
&&W(\vec{r}, \vec{\theta}) = \frac{1}{2\pi \sigma_x \sigma_y} \exp\left( - \frac{x^2}{2\sigma_x^2}\right)
\exp\left( - \frac{y^2}{2\sigma_y^2}\right) \frac{c}{(2\pi)^{4}} \frac{I}{e \hbar} \left(\frac{\omega}{c}\right)^2 \cr && \times \int d \Delta \vec{r} \exp\left(-i \frac{\omega}{c} \vec{\theta} \cdot \Delta
\vec{r}\right)\int d \vec{r}' \widetilde{E}_0(\vec{r}' +\Delta \vec{r}/2) \widetilde{E}_0^*(\vec{r}' - \Delta \vec{r}/2)~.
\label{Wthird}
\end{eqnarray}

Some simplification may be obtained by rewriting the electric field on the source, $\widetilde{E}(\vec{r})$, in the terms of the far field $\widetilde{E}(\vec{\theta})$. In fact, if we write $\widetilde{E}(\vec{r})$ as the integral in Eq. (\ref{farzone}), after substitution in Eq (\ref{Wthird}) we can present results of integration over $\Delta \vec{r}$ and $\vec{r}'$  in terms of the Dirac $\delta$-function and evaluate all integrals analytically. Performing the integration and rearranging yields:

\begin{eqnarray}
W(\vec{r}, \vec{\theta}) = \frac{1}{2\pi \sigma_x \sigma_y} \exp\left(-\frac{x^2}{2\sigma_x^2} -
\frac{y^2}{2\sigma_y^2}\right) \frac{c z_0^2}{(2\pi)^2}  \frac{I}{e \hbar} |\widetilde{E}_0(\vec{\theta})|^2~,
\label{Wthirdfin}
\end{eqnarray}
where $\widetilde{E}_0(\vec{\theta})$ is the field in the far zone generated by a single electron with zero offset and deflection angle. The peak value of the Wigner function is given by

\begin{eqnarray}
B = \max(W) = \frac{1}{2\pi\sigma_x\sigma_y} \max \left(\frac{dF}{d \Omega}\right)~,
\label{Bthirdfin}
\end{eqnarray}
where $\max(dF/d \Omega)$ is the maximum of the angular photon flux from the
wiggler source.

It is relevant to comment on the region of validity of Eq. (\ref{Bthirdfin}). This equation has been derived using the assumption that the transverse size of the electron beam is much larger than maximal size of the intensity distribution at the source in the case of a filament beam. Qualitatively, the wiggler source can be considered as a sequence of periodically spaced bending magnet sources. The characteristic transverse size of the field distribution at a bending magnet source is of order $(R \lambdabar^2)^{1/3}$ , where $\lambdabar=\lambda/(2\pi)$ is the reduced radiation wavelength and $R$ is the bending radius. The radiation from bending magnets always interferes coherently at zero angle with respect to the wiggler axis. This interference is constructive within an angle of about $\sqrt{\lambdabar/L}$, where $L$ is the length of the wiggler. We can estimate the interference size at the source in the middle of wiggler as about $\sqrt{\lambdabar L}$. Obviously, the inequality $\sqrt{\lambdabar L} \gg (R \lambdabar^2)^{1/3}$ is always verified. Additionally, we need to recall the fact that the electron is shifted horizontally by distances $\pm a$ alternately, when it passes through individual wiggler poles. In the wiggler mode operation the wiggling amplitude is larger than the interference size: $a > \sqrt{\lambdabar L}$. In the
opposite case, when $a < \sqrt{\lambdabar L}$ we  deal with the undulator mode of operation. Thus, the requirement for the validity of the beam size dominated-regime can be written as $\sigma_x \gg a$, $\sigma_y \gg \sqrt{\lambdabar L}$. Finally, in the beam size dominated-regime we assume that the electron beam angular divergence is much smaller than the central cone, which is still present in a large-$K$ multipole ideal wiggler due to intrinsic interference effects. This requirement can be written as $\sigma_{x',y'} \ll \sqrt{\lambdabar/L}$.

\section{\label{sec:overview} Overview and earlier results}

It is useful to start our investigation by examining the geometrical properties of a wiggler source. As discussed before, a source placed in the middle of the wiggler is no mathematical abstraction. In fact, one can obtain a visualization of the source by taking a focusing mirror and setting the object plane at the center of the wiggler.

In literature one can often find that the wiggler source properties are described in phase space, that is using geometrical optics as is done in the case of electron beam optics. For example, in the review \cite{WALK} one reads: "The large angular divergence of the radiation emitted by bending magnets and wigglers means that we can ignore diffraction and treat the problem using geometrical optics." In \cite{WALK} this phase space approach is used both in numerical simulations as well as to derive analytical approximated expressions for describing the source and its brightness.

When the electron beam has zero emittance we are dealing with perfectly coherent wavefronts. Intuitively, in this situation one would apply methods from wave optics in order to solve the image formation problem because geometrical optics is intrinsically inadequate to describe the focusing of diffraction limited radiation from a wiggler. Yet, it is possible to use geometrical optics reasoning and obtain an intuitive understanding of the situation. We will describe this approach, and show that such intuitive understanding is in qualitative agreement with an analysis fully based on wave optics. Later on, however, we will also report significant numerical disagreement between exact results and approximated results currently used in literature.

We consider a planar wiggler, so that the transverse velocity of an electron can be written as

\begin{equation}
\vec{v}_\bot(z) = - {c K\over{\gamma}} \sin{\left(k_w z\right)}
\vec{e}_x~, \label{vuzo}
\end{equation}
where $k_w = 2\pi/\lambda_w$ with $\lambda_w$ the undulator period and $K$ the undulator parameter

\begin{equation}
K=\frac{\lambda_w e H_w}{2 \pi m_\mathrm{e} c^2}~, \label{Kpara}
\end{equation}
$m_\mathrm{e}$ being the electron mass and $H_w$ being the maximum of the magnetic field produced by the wiggler on the $z$ axis.

In this case the electron trajectory is given by

\begin{eqnarray}
x(z) = a \cos(k_w z)~,
\label{trajel}
\end{eqnarray}
where $a = K \lambda_w/(2\pi\gamma)$ is the wiggling amplitude.

\begin{figure}[tb]
\includegraphics[width=1.0\textwidth]{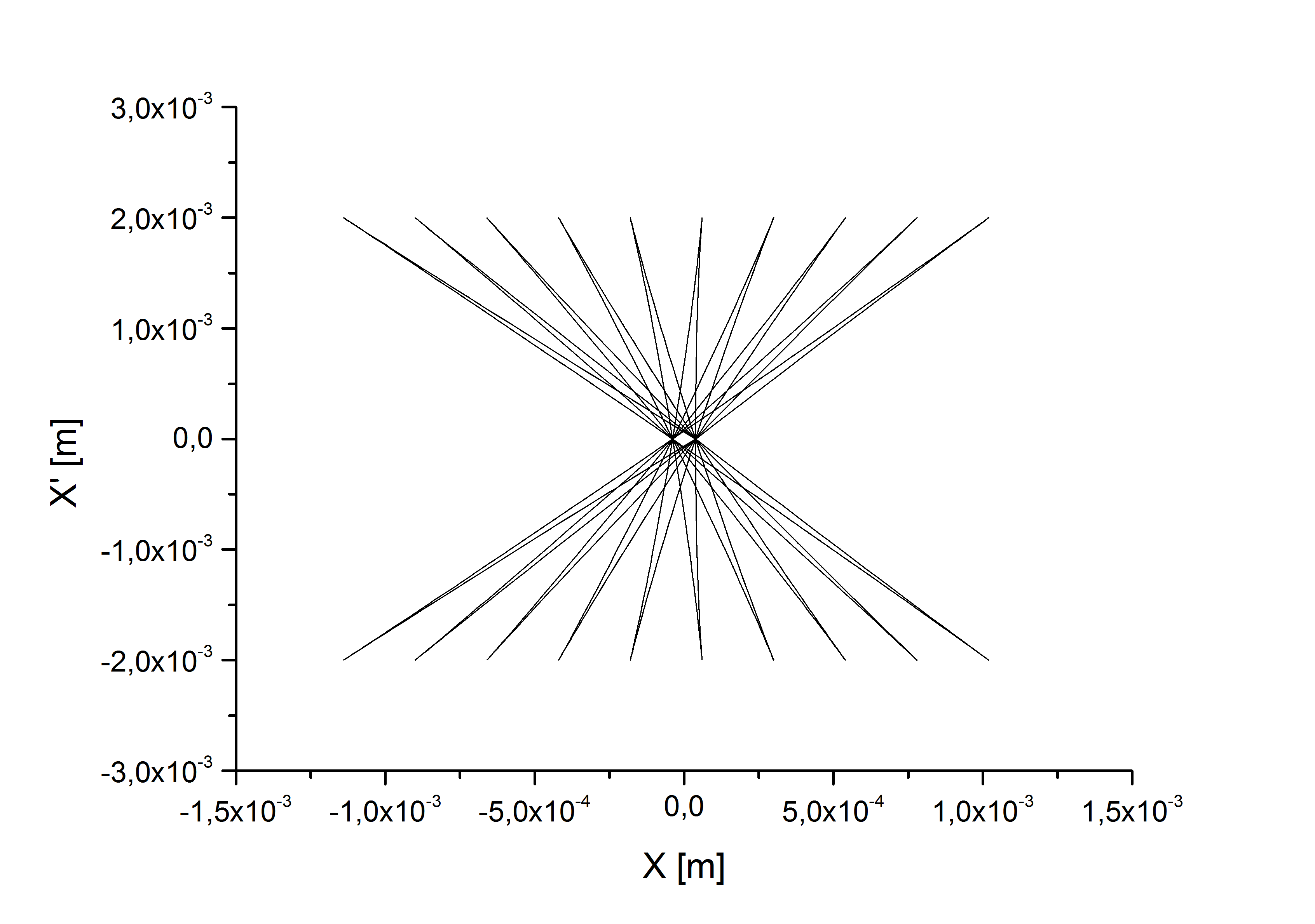}
\caption{The horizontal phase space distribution of radiation at the center of a wiggler with $K = 12$, electron energy $E_{\mathrm{el}} = 3.0$ GeV, $\lambda_w = 12$ cm in the case of $N = 10$ periods according to geometrical optics.} \label{phgeo}
\end{figure}

In the geometrical optics approximation, rays from a single electron are emitted from a given position $z$ in a direction tangent to the trajectory, and are projected backwards or forwards to the reference plane at $z = 0$ that is wiggler center. Fig. \ref{phgeo} shows the horizontal phase space distribution at the wiggler center with $K = 12$, electron energy $E_{\mathrm{el}} = 3.0$ GeV, $\lambda_w = 12$ cm in the case of $N = 10$ periods. We find by inspection $20$ separate strips, each of which corresponds to the magnetic pole distributed along the longitudinal axis, and two source points in the horizontal direction. These two points correspond to the emission from positive and negative poles, separated by a distance $2a$ \cite{WALK}.

Let us consider the apparent horizontal source size as a function of the beamline angular acceptance in the diffraction limited case. The horizontal acceptance angle $\Theta$ indicates an acceptance range $(-\Theta, \Theta)$ that is usually small with respect to the maximum deflection angle $K/\gamma$. In this case, the
emission is restricted to the regions close to each pole, so that flux and bending radius can be considered constant. Since the intensity emitted is independent of $z$, the resulting phase space distribution is
equivalent to  the projection of the tangent to the electron trajectory onto the reference plane at $z = 0$ (see Fig. \ref{phgeo}). In order to obtain an approximate expression for the source size, we first calculate the mean square value $\langle x^2 \rangle$, where brackets $\langle...\rangle$ indicate averaging over the length of the wiggler $(-L/2,L/2)$ and over the acceptance range $(-\Theta, \Theta)$. By assuming $N \gg 1$ the following approximation can be obtained for an electron beam with zero emittance:

\begin{eqnarray}
\langle x^2 \rangle = \langle (a+z x')^2 \rangle = a^2 + \frac{L^2 \Theta^2}{36}~.
\label{msv}
\end{eqnarray}
We can now account for a finite phase space distribution for the electron beam. The resulting photon phase space distribution at $z =0$ is equivalent to a convolution of the ellipse describing the electron beam phase-space, with the horizontal phase space distribution at $z = 0$ that pertains the emission from a single electron (see Fig.  \ref{phgeo}). Following \cite{WALK}, the middle of the wiggler at $z = 0$ is taken to be a symmetry point of the electron beam lattice, and the electron beam distribution can be written as

\begin{eqnarray}
f(x,x') = \frac{1}{2 \pi \sigma_x \sigma_{x'}} \exp\left(- \frac{x^2}{2\sigma_x^2}\right) \exp\left(- \frac{x'^2}{2\sigma_x'^2}\right)~,
\label{elphasp}
\end{eqnarray}
where $\sigma_x$ is the rms horizontal electron beam size at $z = 0$ and $\sigma_{x'}$ is the rms horizontal electron beam divergence. The effective source size is found by adding $\sigma_x$ and the diffraction limited size in Eq. (\ref{msv}) in quadrature, thus obtaining

\begin{eqnarray}
\Sigma_x = \left[\sigma_x^2 + a^2 + \frac{L^2 \Theta^2}{36} \right]^{1/2}~.
\label{Sigtot}
\end{eqnarray}
It is easy to see that for any shift along the $x'$ direction of the distribution in Fig.\ref{phgeo}  the projection on the $x$ axis remains the same. Therefore, the source size is insensitive to the angular  distribution of the electron beam.

The results presented above for the wiggler source are in contrast with what is reported in literature. For example, the effective horizontal source size of a wiggler radiation source is given in Eq. (26) of reference \cite{WALK} as:

\begin{eqnarray}
\Sigma_x = \left[\sigma_x^2 +a^2 + \frac{L^2 \sigma_{x'}^2}{12} + \frac{L^2 \Theta^2}{36}\right]^{1/2}~,
\label{Sigtotlit}
\end{eqnarray}
where the third term is the result of the so-called depth-of-field effects, contributions to the apparent source size from different poles. From beam dynamics considerations it is known that, if focusing elements
are absent, the beam size varies along the wiggler like

\begin{eqnarray}
\sigma_x^2(z) = \sigma_x(0)^2 + z^2 \sigma_{x'}^2 ~,
\label{sigel2}
\end{eqnarray}
where $-L/2 < z < L/2$. It follows that the average beam size along the wiggler length is

\begin{eqnarray}
\langle \sigma_x^2 \rangle = \sigma_x^2(0) + L^2 \frac{\sigma_{x'}^2}{12}
\label{avesigel2}
\end{eqnarray}
In papers \cite{HUL1, HUL2} and review \cite{WALK},  Eq. (\ref{avesigel2}) is regarded as a clear evidence of the fact that the effective source size is widened by depth-of-fields effects. However this is misconception, because according to the phase space approach the source size is not widened at all.

The criticism we just expressed is focused on the third term in Eq. (\ref{Sigtotlit}), which describes the source size widening due to depth-of-field effects and should not be there. There is another objection that could be made to the analysis \cite{WALK}, related with fourth term in Eq. (\ref{Sigtotlit}), which concerns quantitative aspects of the dependence of the effective source size on the horizontal beamline aperture. According to Eq. (\ref{Sigtot}) the effective source size is widened in proportion to the length of the wiggler $L$ and to the beamline opening angle $\Theta$. After converting rms values to FWHM, assuming a Gaussian form for the effective source size, the behavior of the Eq. (\ref{Sigtot}) for the NSLS-II damping wiggler was analyzed  in \cite{BERM}. It was found that effect of "blurring of the effective horizontal source size in this case will be dramatic due to very long ($L = 7$ m) wiggler length". In the next section, however, we will demonstrate that the quantitative agreement between the approximation in Eq. (\ref{Sigtot}) and exact results is rather poor. In particular, in realistic situations the approximation in Eq. (\ref{Sigtot}) overestimates the exact value of the source FWHM by an order of magnitude.

One concludes that the application of geometrical optics reasoning yield only an intuitive understanding of the situation, but quantitative studies should be undertaken within a wave-optics framework. First-principle computer codes (see e.g. \cite{CHUB} -\cite{SPECTRA}) have been  successfully used to model advanced SR sources and beamlines without specific analytical simplifications. Results may be obtained using numerical techniques alone, starting from the Lienard-Wiechert expression for the electromagnetic field and using only the ultra-relativistic approximation. Similarly, these codes also allow for a treatment of the problem of imaging a wiggler source. It is therefore instructive to reconsider the problem of the prediction of the size of a wiggler source  by means of numerical simulations alone, which play the same role of an  experiment. In the following of this article we will consider two cases at fixed electron beam size:

\begin{itemize}
\item The usual case with matched beta function $\beta_0 \simeq L$ and additionally $\sigma_x^2 = \epsilon_0\beta_0$ and  $\sigma_{x'}^2 = \epsilon_0/\beta_0$ .

\item A case with tenfold increase in emittance and tenfold decrease in beta function, i.e. $\epsilon = 10 \epsilon_0$, and $\beta = \beta_0/10$ .
\end{itemize}

Of course one is free to choose other numerical cases. In section \ref{sec:sims} we will show results of simulations confirming our prediction that the image of the source is insensitive to the electron beam divergence for the two case above. This is in contrast with the prediction\cite{HUL1,HUL2,WALK} of a significant source widening in the second numerical case compared to the first.

Note that according to literature, photons in wiggler are emitted incoherently in the tangential direction at each point of the electron trajectory. It should be clear that one can talk about incoherently emitted photons only in the framework of statistical optics, when one deals with SR as a random process. In the case of a single electron we are always dealing with coherently emitted photons at each point of the trajectory. In fact, the entire photon flux collected from a diffraction limited source is fully transversely coherent. A Young's double pinhole interferometer can be used for demonstrating this fact. In the case of a filament electron beam (i.e. a beam with zero emittance), the interference pattern recorded by the interferometer is always characterized by a $100 \%$ fringe contrast, which is in fact defined as  the modulus of the degree of transverse coherence. Intuitively, in order to solve the image formation problem in this situation, one would apply methods from wave optics. At variance here we discussed an estimation of the source size for the case of diffraction limited radiation from a wiggler, based on geometrical optics, as is done in literature. A situation analogous to the one we have just examined is the calculation of a laser beam focus through a lens, when severe aberrations are present. Although the laser beam is coherent, when diffraction effects are negligible compared to aberration effects the beam focusing can be calculated with the help of geometrical optics considerations alone \cite{BORN}. The similarity between the two situations is highlighted by the essential feature of diffraction limited SR beam from bending magnet: at a horizontal angle $\theta_x$ larger than diffraction limited vertical opening angle $(\lambda/R)^{1/3}$, the wavefront aberrations are present in the sense discussed above, and are severe \cite{OURJSR}.  Therefore, a geometrical optics approximation, leading in particular to ray-tracing techniques, can be applied to the analysis of the wiggler image formation problem. As a result, for the single electron in Fig. \ref{phgeo} we deal with a `pseudo' phase space, where the separation of the linear stripes from one another  has no physical sense.

We emphasize here an important distinction between our application of the phase space method  and  that in literature, where the depth-of-field effect erroneously appears. Such distinction is in the order of execution of two distinct operations: the averaging over the electron beam phase space distribution and the summation over wiggler poles ( i.e. the linear ridges in Fig. \ref{phgeo} ), which in the limit for $N \gg 1$ can be approximated  by integrating along the wiggler length. The point is that these operations do not commute. Therefore, our result is in contrast with that presented in literature. As we will demonstrate by numerical analysis in the next section, our analysis is the correct one. So here we must not average over the beam phase space separately over different ridges. We first need to sum over all ridges and then average over an ensemble of electrons.

Finally, we should make a few remarks concerning the terminology used here in relation to the treatment of single electron radiation. In fact, one can see a net distinction between the phase space method, when one discusses about averages over an ensemble of electrons, and the geometrical optics limit in the framework of coherent optics, when one discusses about a highly aberrated beam radiated from a single electron in the wiggler setup. We used, as in literature, the wording `mean square value' and `averaged over the wiggler length' in a diffraction limited case. Actually, the quantities discussed in such situations must not be understood as averaged over some ensemble, but as the result of the application of the geometrical optics approximation when aberration effects are dominant compared to diffraction effects.

We now turn to the main topic of this study, namely the analysis of the brightness of wiggler sources. An usual estimate, proposed in \cite{WALK}, is given by

\begin{eqnarray}
B = N_{\mathrm{pol}} \frac{dF}{d \theta_x} \frac{1}{(2\pi)^{3/2} \Sigma_x \Sigma_y \Sigma_y'}~ ,
\label{B1}
\end{eqnarray}
where $N_{\mathrm{pol}} = 2N$ is the number of wiggler poles and $dF/d \theta_x$ is the spectral photon flux per unit horizontal angle from a single pole. The effective horizontal and vertical source size and effective vertical
divergence are calculated as

\begin{eqnarray}
&& \Sigma_x    = \left[\sigma_x^2 +a^2 + \frac{L^2}{12}\sigma_{x'}^2\right]^{1/2}~   ,\cr
&& \Sigma_y    = \left[\sigma_y^2      + \frac{L^2}{12}\sigma_{y'}^2\right]^{1/2}~   ,\cr
&& \Sigma_{y'} = \left[\sigma_{y'}^2   + \sigma_{r'}^2\right]^{1/2}~ ,
\label{sigs}
\end{eqnarray}
where the vertical opening angle $\sigma_{r'}$ can be determined from the equality

\begin{eqnarray}
(2\pi)^{1/2} \sigma_{r'}\left( \frac{dF}{d \Omega}\right)_{\theta_{y} = 0}  = \frac{dF}{d \theta_{x}}~,
\label{verang}
\end{eqnarray}
where $(dF/ d\Omega)_{\theta_{y} = 0}$ is the on-axis spectral  photon flux density per unit solid angle radiated from a single pole.

The most serious objection to approximation in Eq. (\ref{B1}) is that this expression accounts for the widening of the effective source sizes $\Sigma_x$ and $\Sigma_y$  due to `depth-of-field' effects which, as we discussed before, does not exist.

Another argument that disqualifies Eq. (\ref{B1}) as a good approximation for the wiggler brightness follows from the comparison with exact results found in the previous section. In the beam size-dominated regime when
$\sigma_x^2 \gg a^2$, $\sigma_{x'}^2 \ll \sigma_{r'}^2$, $\sigma_{y'}^2 \ll \sigma_{r'}^2$, and after deleting the depth-of-field terms,  Eq. (\ref{B1}) yields

\begin{eqnarray}
B = \frac{N_\mathrm{pol}}{2\pi \sigma_x \sigma_y} \left(\frac{dF}{d \Omega}\right)_{\theta_y=0}~.
\label{B2}
\end{eqnarray}
However, we can find the brightness corresponding to the same limiting situation using a rigorous mathematical method, that is following the Wigner function approach. Then, from our definition of brightness in the previous section we find (see Eq. (\ref{Wcorr}))

\begin{eqnarray}
B = \frac{1}{2\pi\sigma_x \sigma_y}\max\left(\frac{dF}{d \Omega}\right)~  ,
\label{B3}
\end{eqnarray}
where $dF/d \Omega$ is the angular spectral flux density of the radiation from a wiggler. It is evident that the maximum of the angular spectral flux density generated by a wiggler in the approximate Eq. (\ref{B3}) is  replaced by the number of poles multiplied by the maximum flux density from a single pole. In other words, any interference effect is neglected. In practical situations such assumption can lead to significant underestimation of the brightness. As we will demonstrate in the following section, $dF/d \Omega$ is a highly oscillatory function, and its maximum strongly depends on wiggler field errors. In particular it can be orders of magnitude higher than in the estimation used in Eq. (\ref{B2}).  The best way to avoid this kind of difficulties, which can be used routinely by SR beamline scientists, is to use codes like SRW for calculating $\max(dF/d \Omega)$, as we will demonstrate in section \ref{sec:sims}.

\section{\label{sec:sims} Simulations of wiggler source properties}

In this section we compare the geometrical properties of the wiggler source obtained by numerical simulations with estimates obtained from phase space analysis. We performed simulations using the code SRW \cite{CHUB}. The goal of our simulations is to determine the exact radiation output from a wiggler, and the geometrical properties of the virtual source located in the middle of the wiggler.

To compute SR properties, many parameters concerning the electron beam and wiggler setup should be specified. This is done in Table \ref{tab1} where, $\lambda_w$ is the period length, $B_w$ is the peak magnetic field, $K$ is the undulator parameter given by $K=0.934  B_w[T] \lambda_w[cm]$ and $N$ is the number of periods. The emitted radiation is considered at a distance of $10$ m from the wiggler center. In all cases SRW calculates SR properties in the near-field region, subject only to the paraxial approximation.

\begin{table}
\caption{Parameters of electron beam and wiggler setup used in simulations.}

\begin{small}\begin{tabular}{ l c c}
\hline           & ~ Units           &  ~    \\ \hline
N                & -                 & 30    \\
$\lambda_w$      & cm                & 12    \\
$B_w$            & T                 & 1.097 \\
K                & -                 & 12.3  \\
Electron energy  & GeV               & 3.0   \\
Beam current     & A                 & 0.275 \\
Photon energy    & eV                & 938   \\
\hline
\end{tabular}\end{small}
\label{tab1}
\end{table}

The field calculated consists of two components, referring to the two directions of polarization, horizontal and vertical. The radiation intensity presented in figures is the total one instead, and is obtained by summing over the two polarization contributions.

\subsection{Ideal wiggler}

\begin{figure}[tb]
\includegraphics[width=1.0\textwidth]{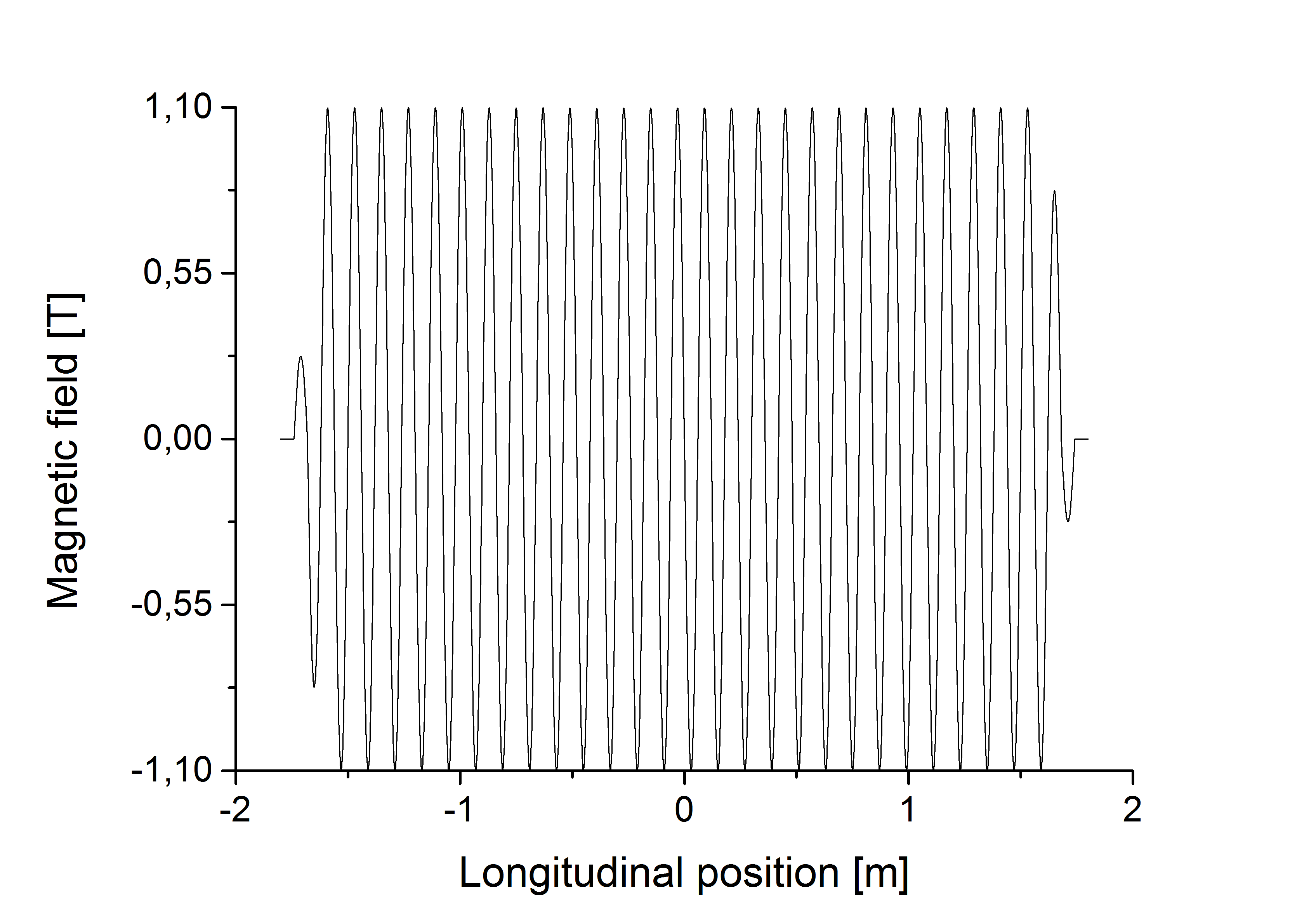}
\caption{Magnetic field for an ideal planar wiggler} \label{fieldfig}
\end{figure}

\begin{figure}[tb]
\includegraphics[width=1.0\textwidth]{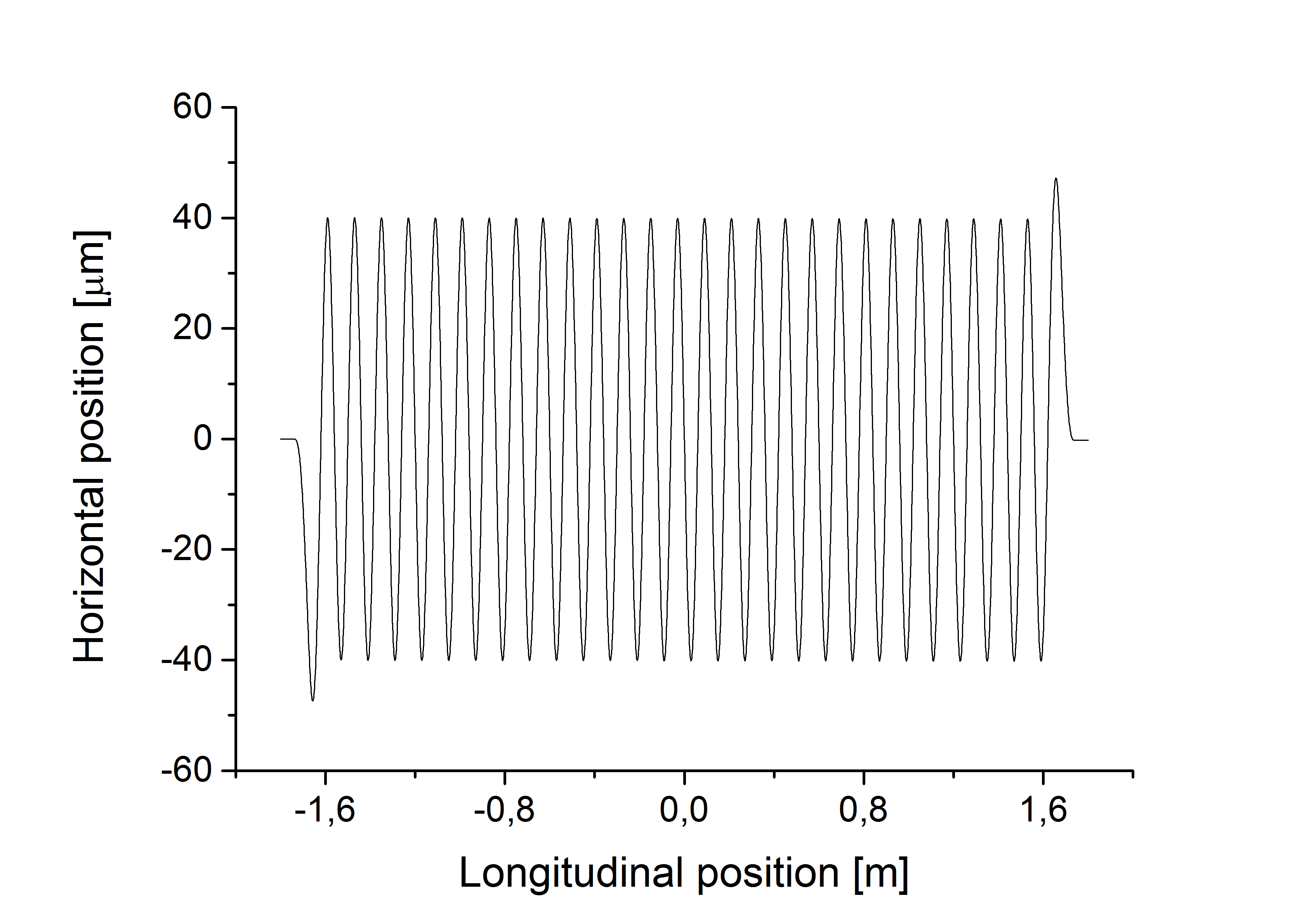}
\caption{Trajectory of the 3 GeV reference electron passing through a wiggler with magnetic field levels shown in Fig. \ref{trajfig} } \label{trajfig}
\end{figure}

\begin{figure}[tb]
\includegraphics[width=1.0\textwidth]{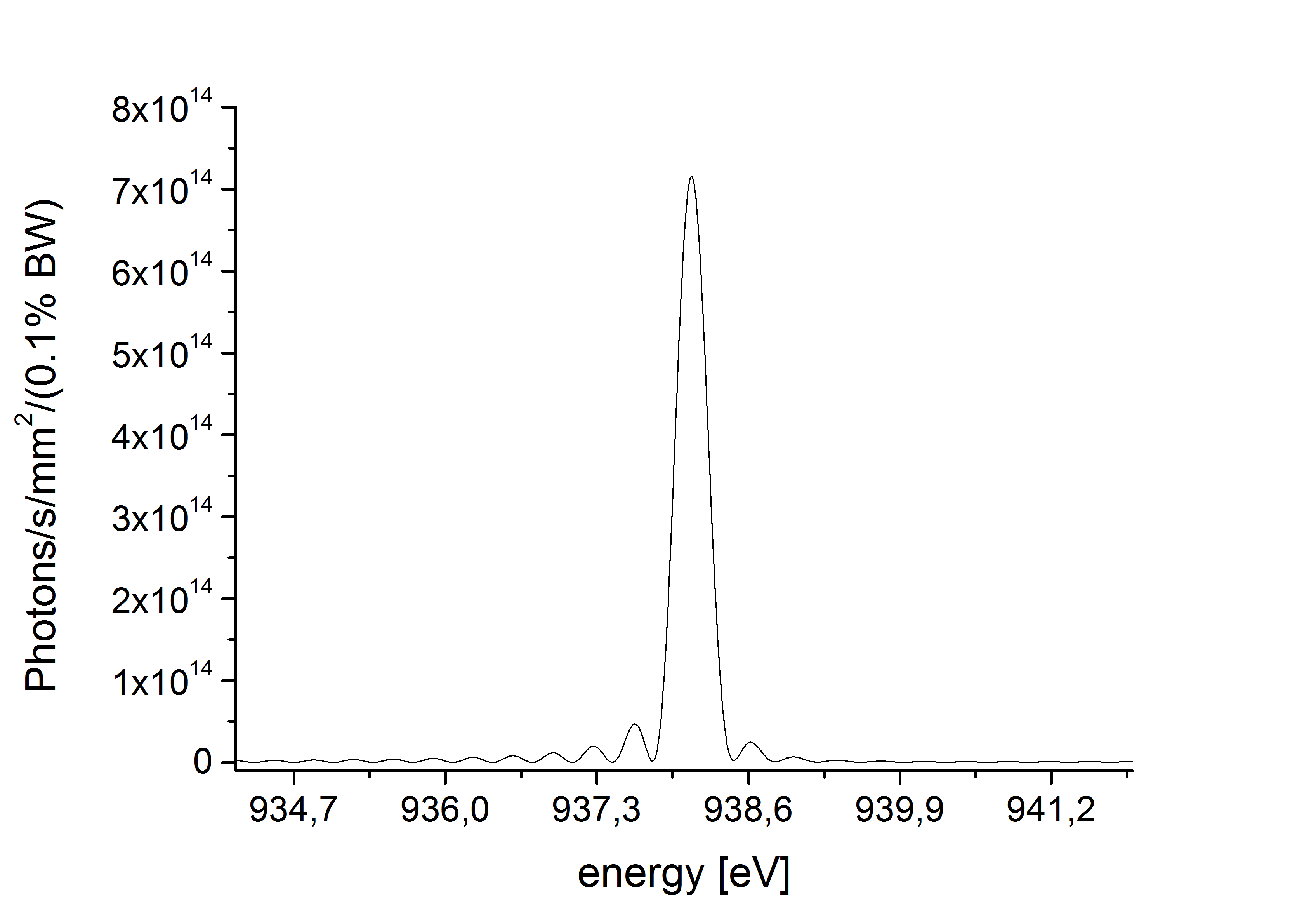}
\caption{Ideal wiggler. Spectrum through a centered slit (with dimensions $0.1$ mm by $0.1$ mm) placed at $10$ m from the middle of the wiggler. The flux density is plotted as a function of the photon energy around the 101st harmonic for an electron beam with zero emittance and energy spread. } \label{specid}
\end{figure}

\begin{figure}[tb]
\includegraphics[width=1.0\textwidth]{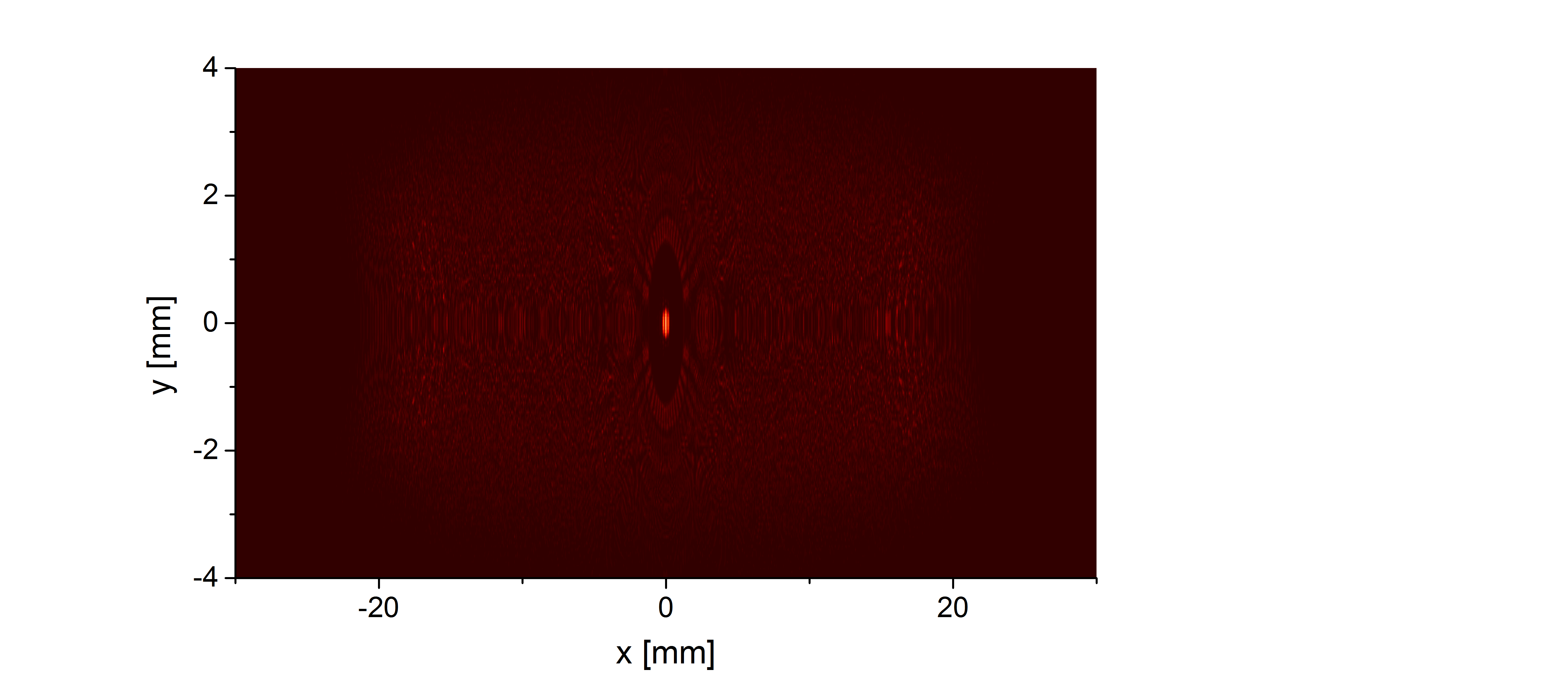}
\caption{Ideal wiggler. Transverse flux distribution at 10 m from the middle of the wiggler  for an  electron beam with zero emittance and energy spread. } \label{fluxideal}
\end{figure}

\begin{figure}[tb]
\includegraphics[width=1.0\textwidth]{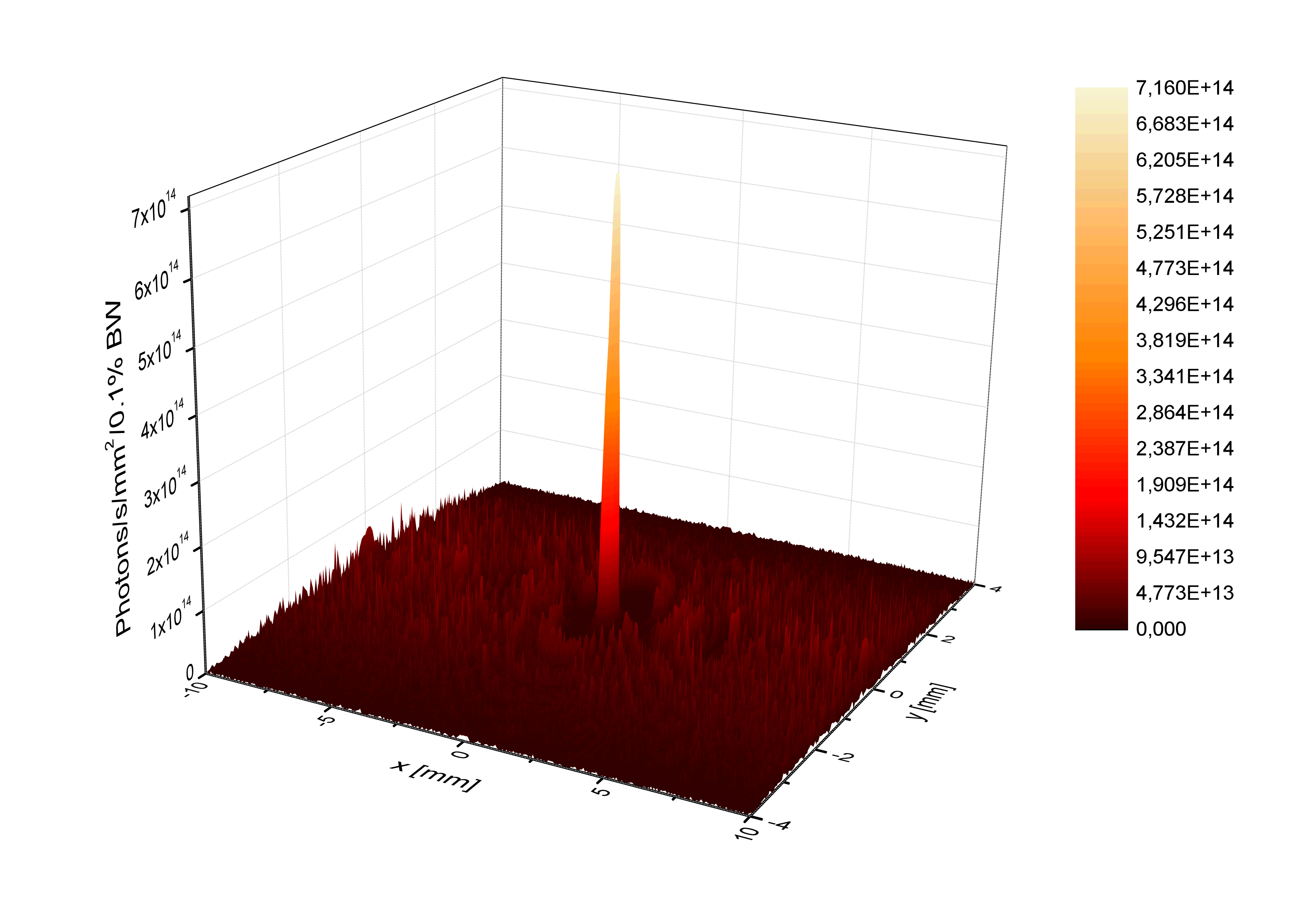}
\caption{Ideal wiggler. 3D view of  Fig. \ref{fluxideal} for a different coordinates range. } \label{fluxideal_3D}
\end{figure}

\begin{figure}[tb]
\includegraphics[width=1.0\textwidth]{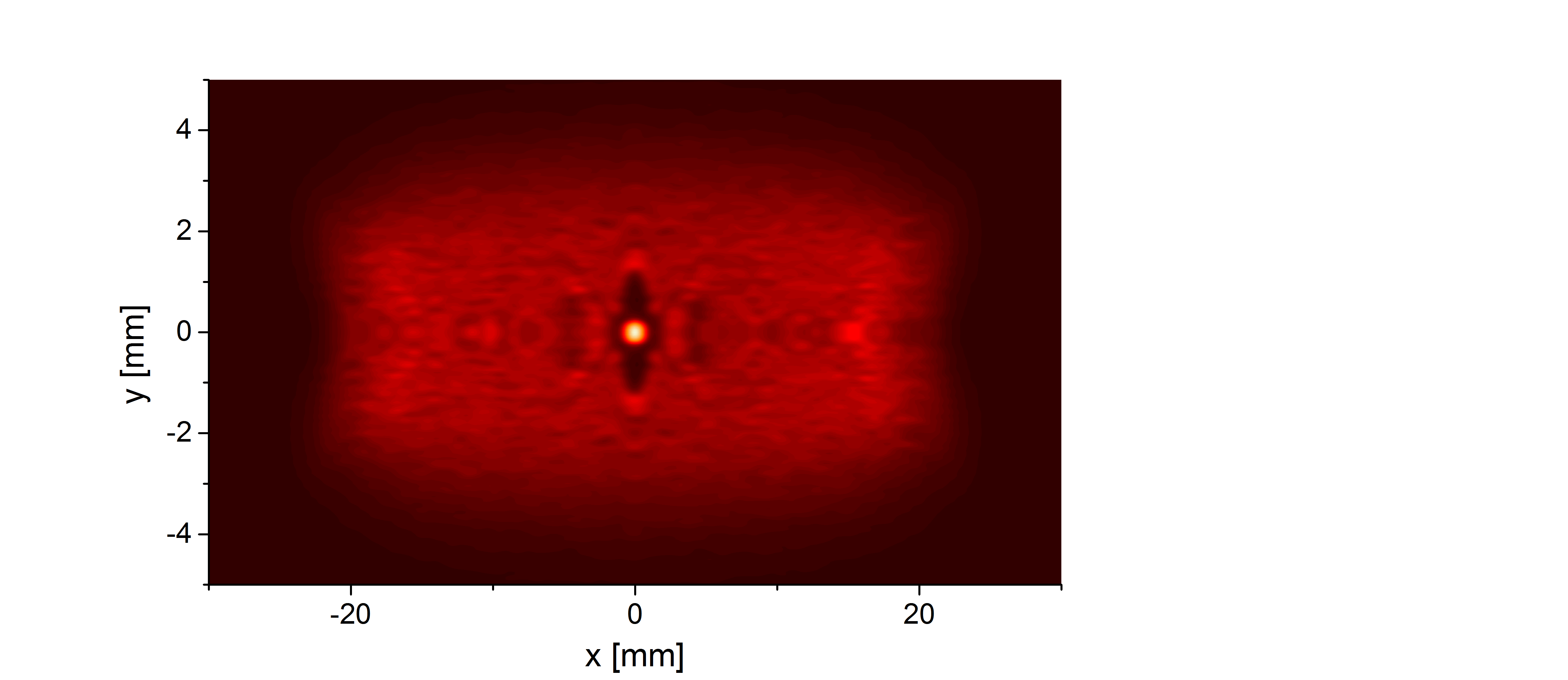}
\caption{Ideal wiggler. Transverse flux distribution at 10 m from the middle of the wiggler  with horizontal emittance $\epsilon_x = 1$ nm, horizontal beta function $\beta_x = 1$ m,  vertical emittance $\epsilon_y = 0.01$ nm, vertical beta function $\beta_y = 1$ m.} \label{fluxemit}
\end{figure}

\begin{figure}[tb]
\includegraphics[width=1.0\textwidth]{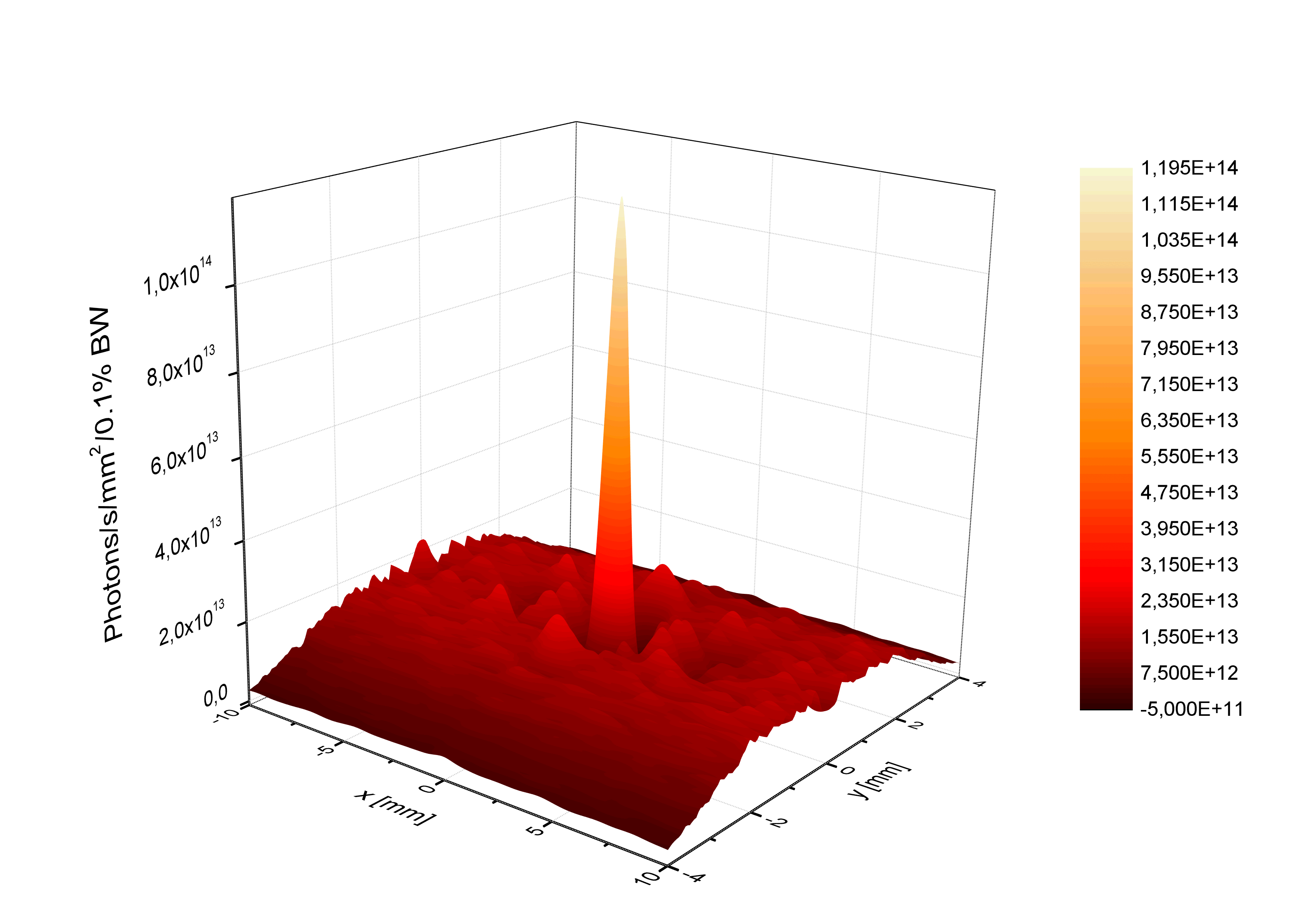}
\caption{Ideal wiggler. 3D view of Fig. \ref{fluxemit} for a different coordinates range.  } \label{fluxemit_3D}
\end{figure}

\begin{figure}[tb]
\includegraphics[width=1.0\textwidth]{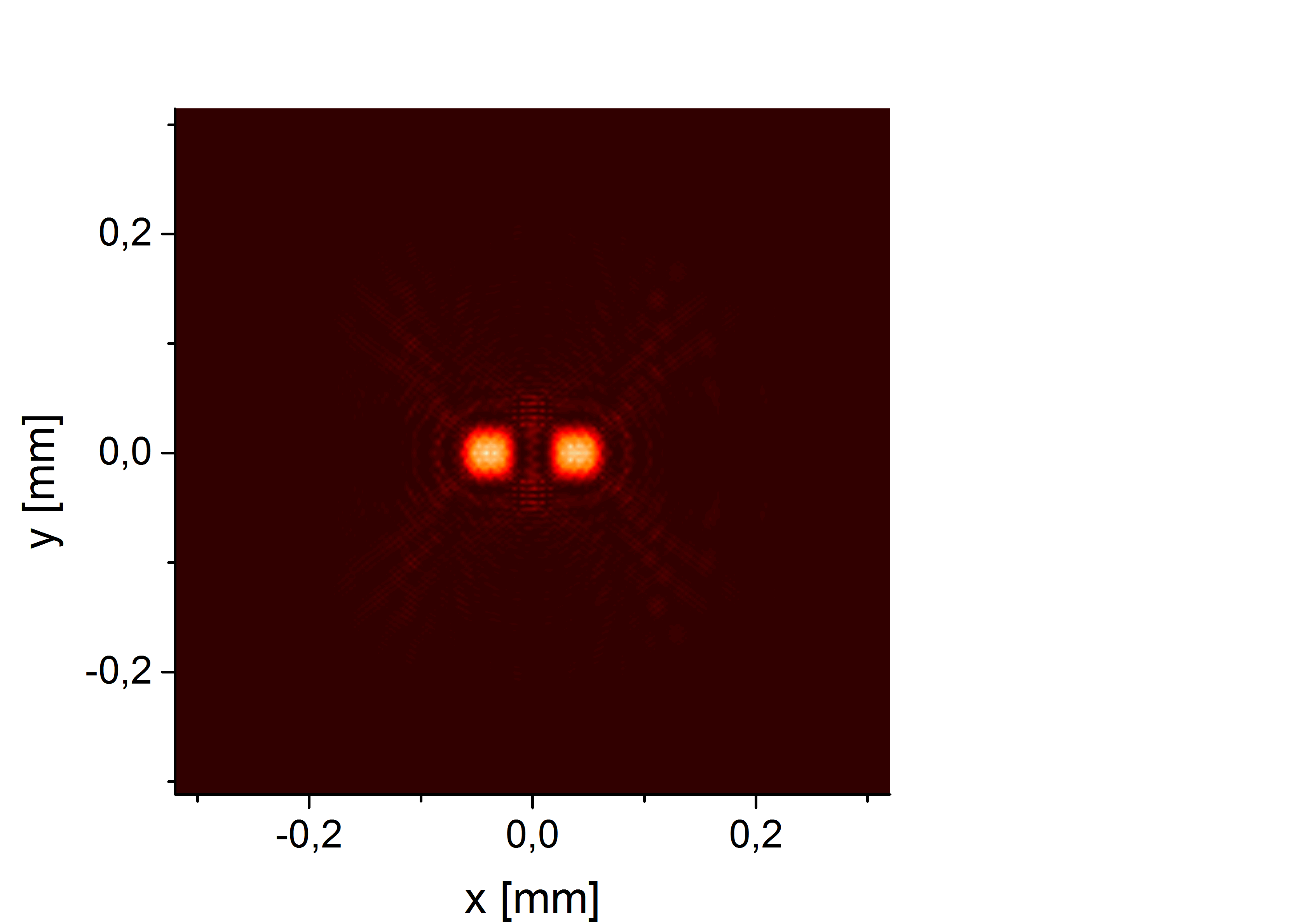}
\caption{Ideal wiggler. Intensity distribution at the virtual source placed in the middle of the wiggler without electron beam emittance and energy spread. The image is obtained simulating a perfect lens immediately behind a centered rectangular aperture of $2.5$ mm by $2.5$ mm, placed at $10$ m from the middle of the wiggler.} \label{intevirt1}
\end{figure}

\begin{figure}[tb]
\includegraphics[width=1.0\textwidth]{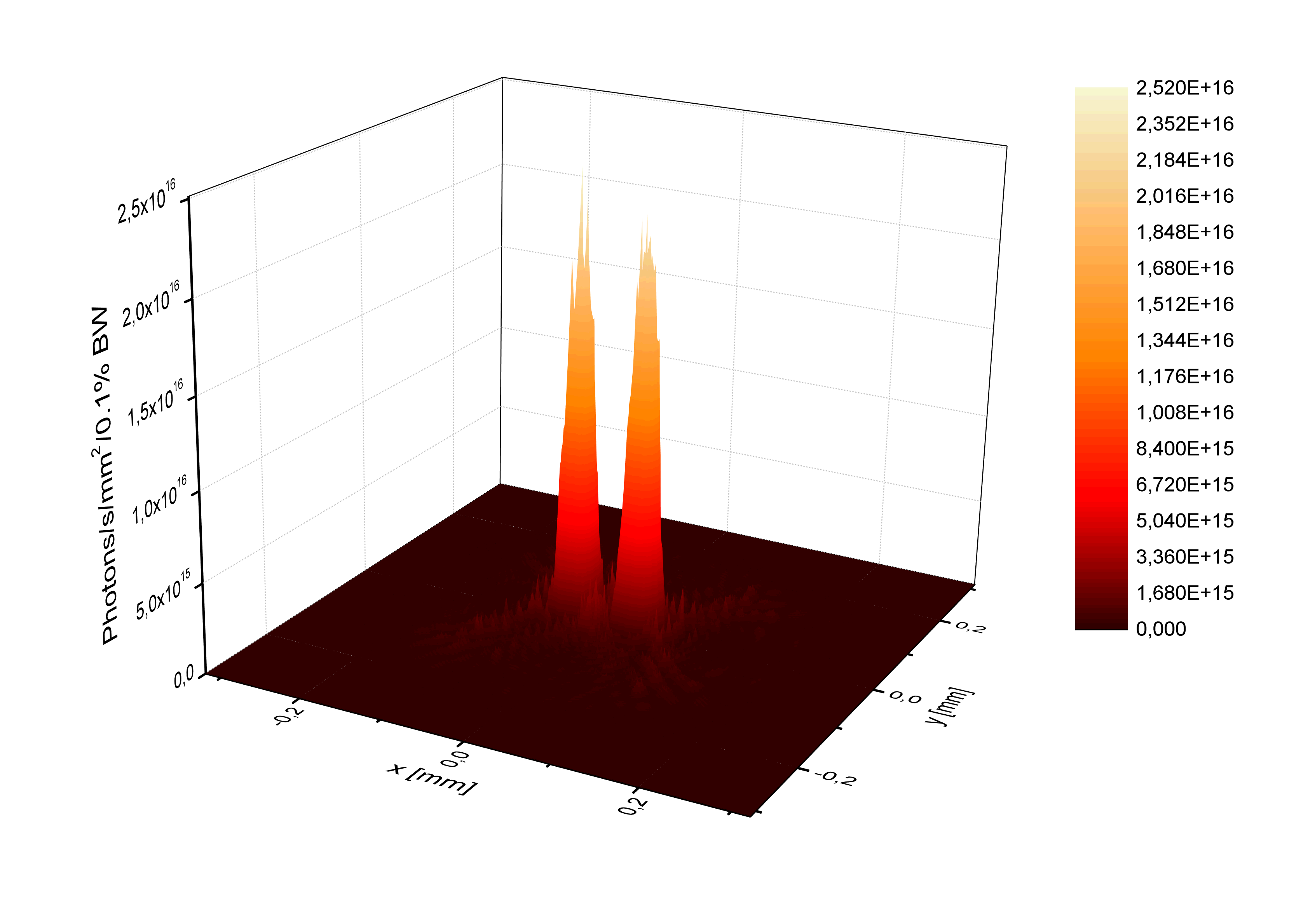}
\caption{Ideal wiggler. 3D view of Fig. \ref{intevirt1}, but for a different coordinate range.} \label{intevirt1_3D}
\end{figure}

\begin{figure}[tb]
\includegraphics[width=0.5\textwidth]{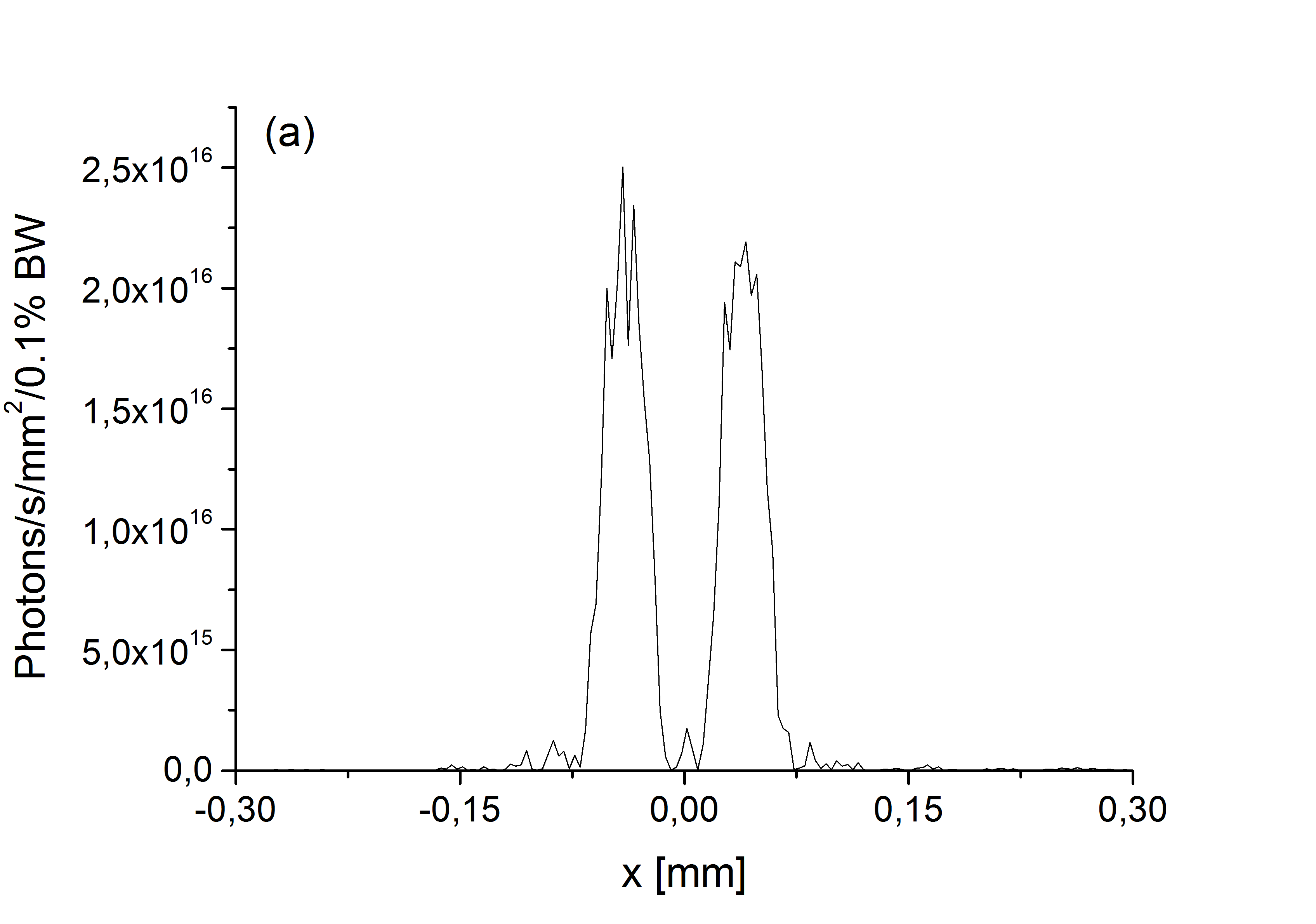}
\includegraphics[width=0.5\textwidth]{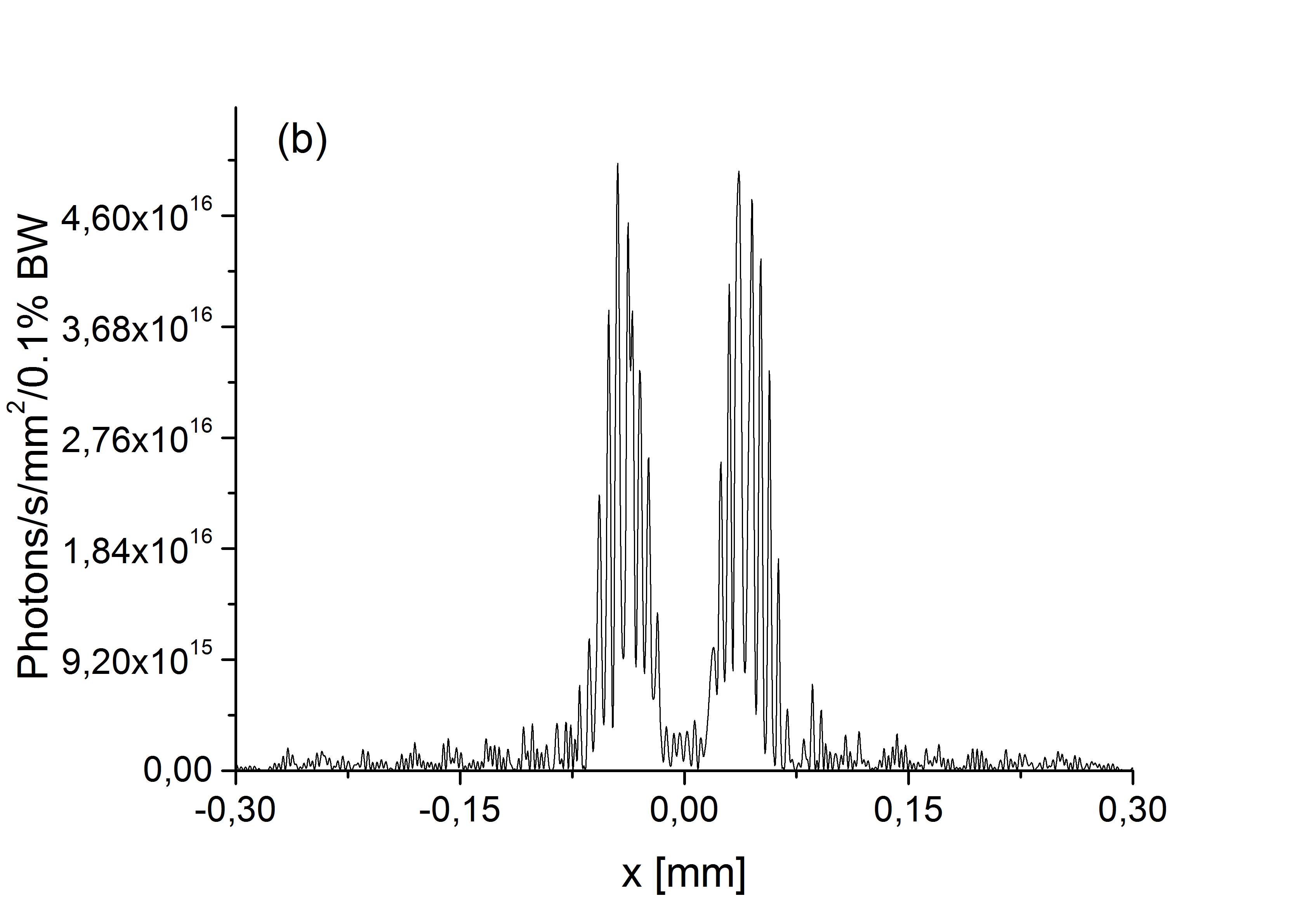}
\caption{Ideal wiggler. Horizontal cut at the median plane of the intensity profile at the source. The image is obtained simulating a perfect lens immediately behind a centered rectangular aperture  with varying horizontal size at $10$ m from the middle of the  wiggler without the inclusion of electron beam emittance and energy spread. The slit considered here is in $2.5$ mm wide along the vertical direction. (a) The slit width is 2.5 mm along the horizontal direction. (b) The slit width is 5 mm along the horizontal direction. 
} \label{slitvar}
\end{figure}

\begin{figure}[tb]
\includegraphics[width=1.0\textwidth]{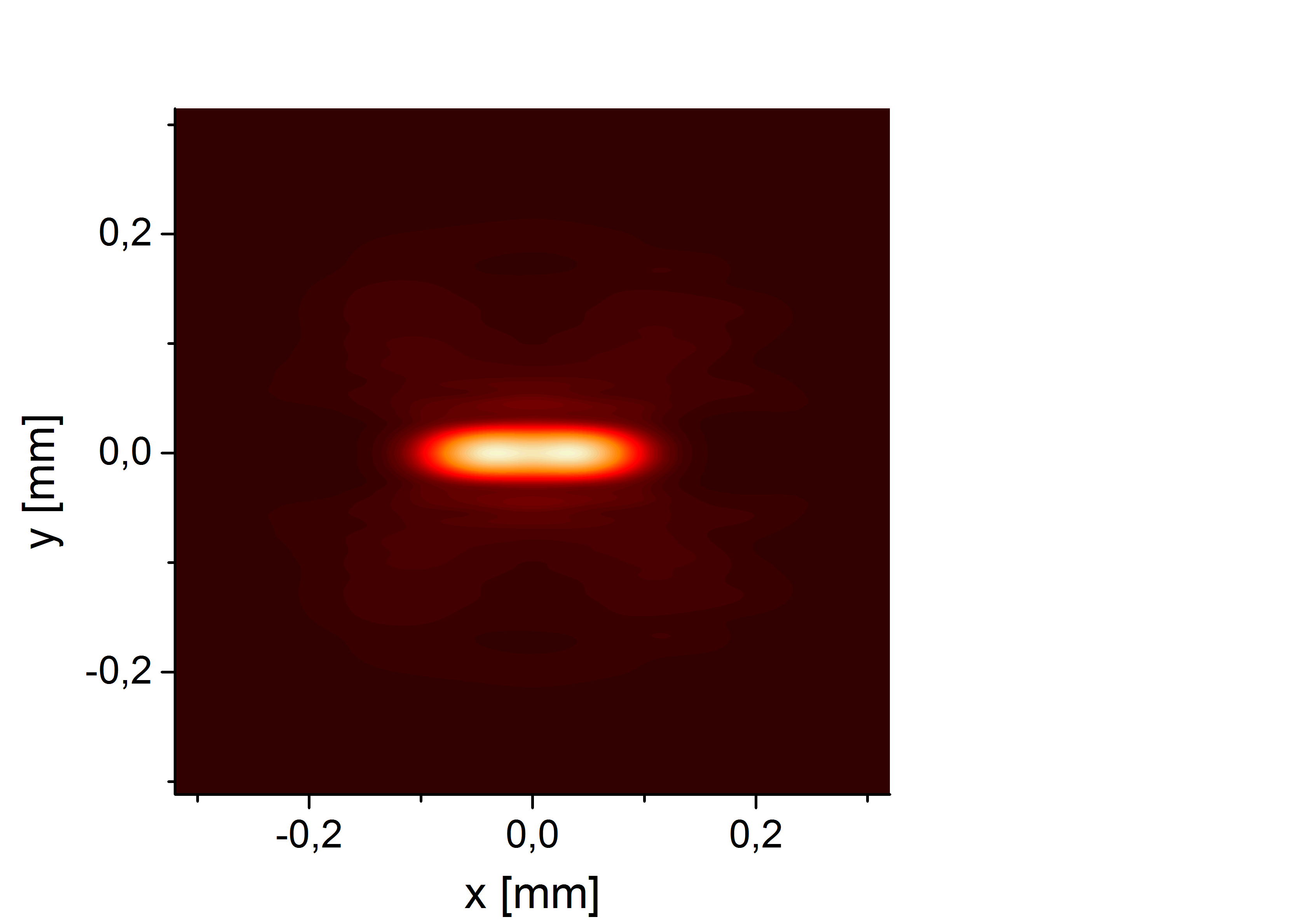}
\caption{Ideal wiggler. Intensity distribution at the virtual source placed in the middle of the wiggler. Here $\epsilon_x = 1$ nm and $\beta_x = 1$ m. In the vertical direction $\beta_y = \beta_x$, and $\epsilon_y = \epsilon_x/100$. The image is obtained simulating a perfect lens immediately behind a centered rectangular aperture of $2.5$ mm by $2.5$ mm, placed at $10$ m from the middle of the wiggler.
} \label{intevirt_emit}
\end{figure}

\begin{figure}[tb]
\includegraphics[width=1.0\textwidth]{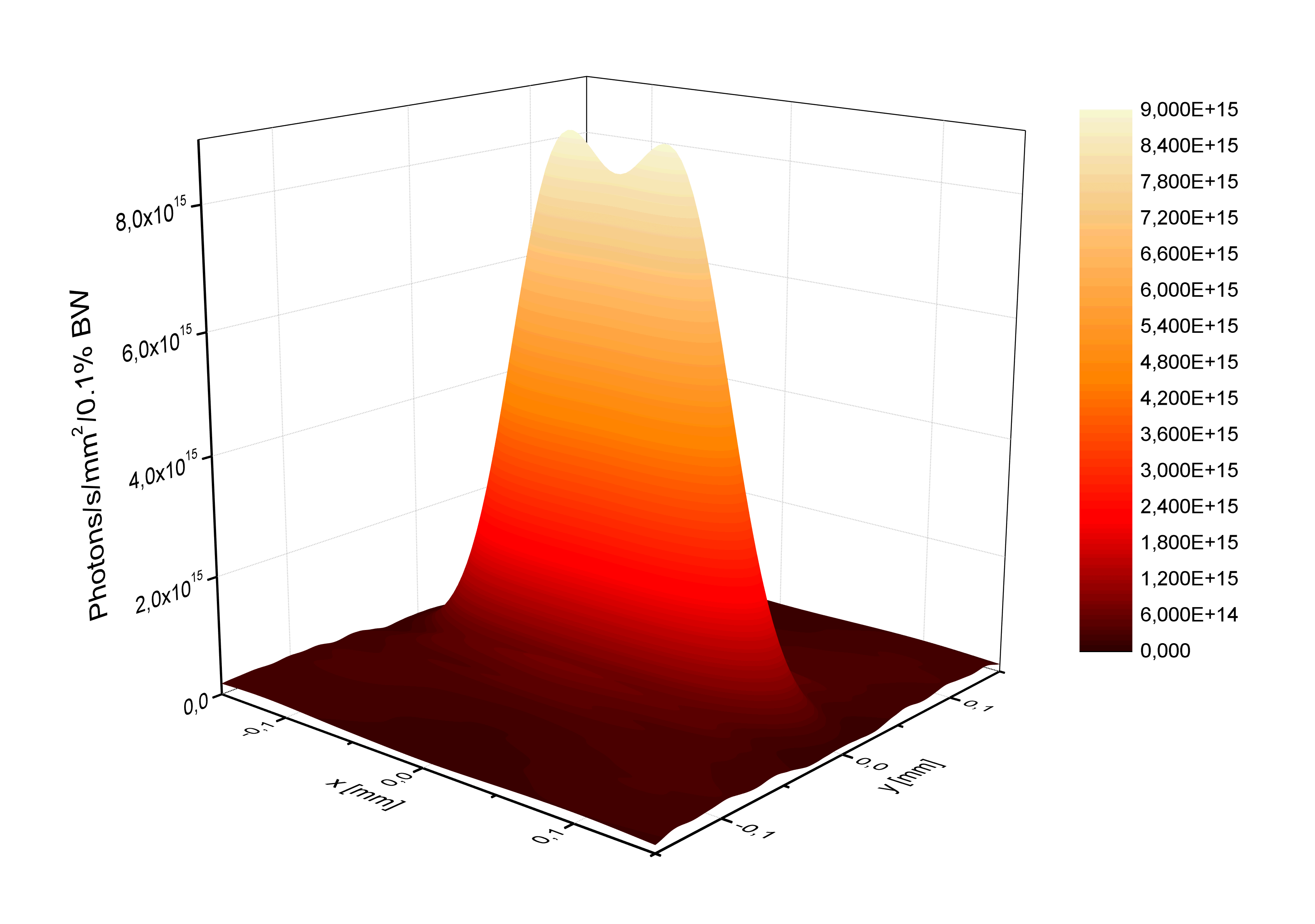}
\caption{Ideal wiggler. 3D view of the Fig. \ref{intevirt_emit}, but for a different coordinate range.} \label{intevirt_emit_3D}
\end{figure}

\begin{figure}[tb]
\includegraphics[width=1.0\textwidth]{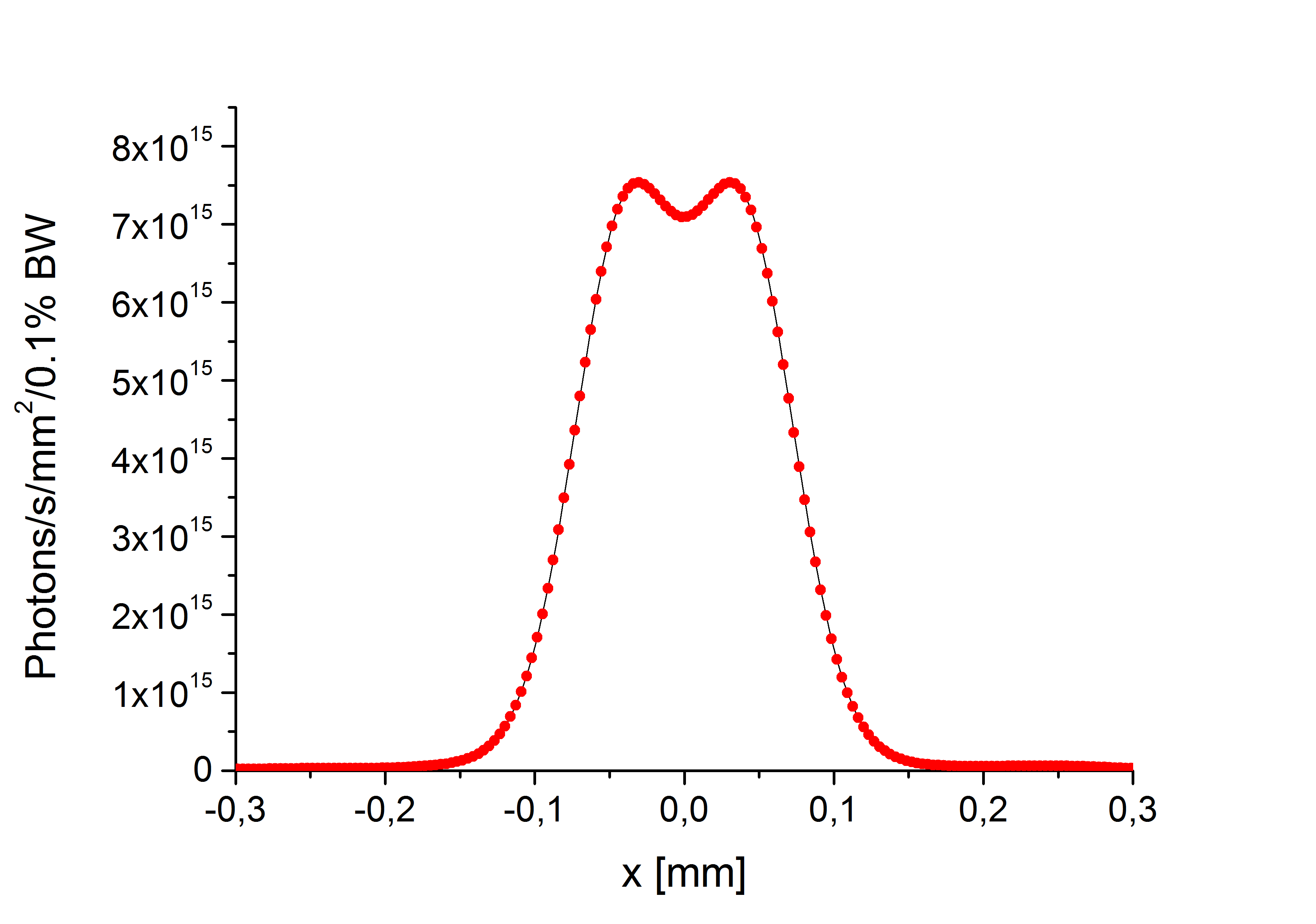}
\caption{Ideal wiggler. Horizontal cut at the median plane  of the intensity profile at the source. The image is obtained simulating a perfect lens immediately behind a centered rectangular aperture of $2.5$ mm by $2.5$ mm, placed at $10$ m from the middle of the wiggler. The solid line corresponds to $\epsilon_x = 1$ nm, $\beta_x = 1$ m, while the circles correspond to $\epsilon_x = 10$ nm, $\beta_x = 0.1$ m. In the vertical direction, for both cases, $\beta_y = \beta_x$, and $\epsilon_y = \epsilon_x/100$.} \label{horcutid}
\end{figure}

\begin{figure}[tb]
\includegraphics[width=1.0\textwidth]{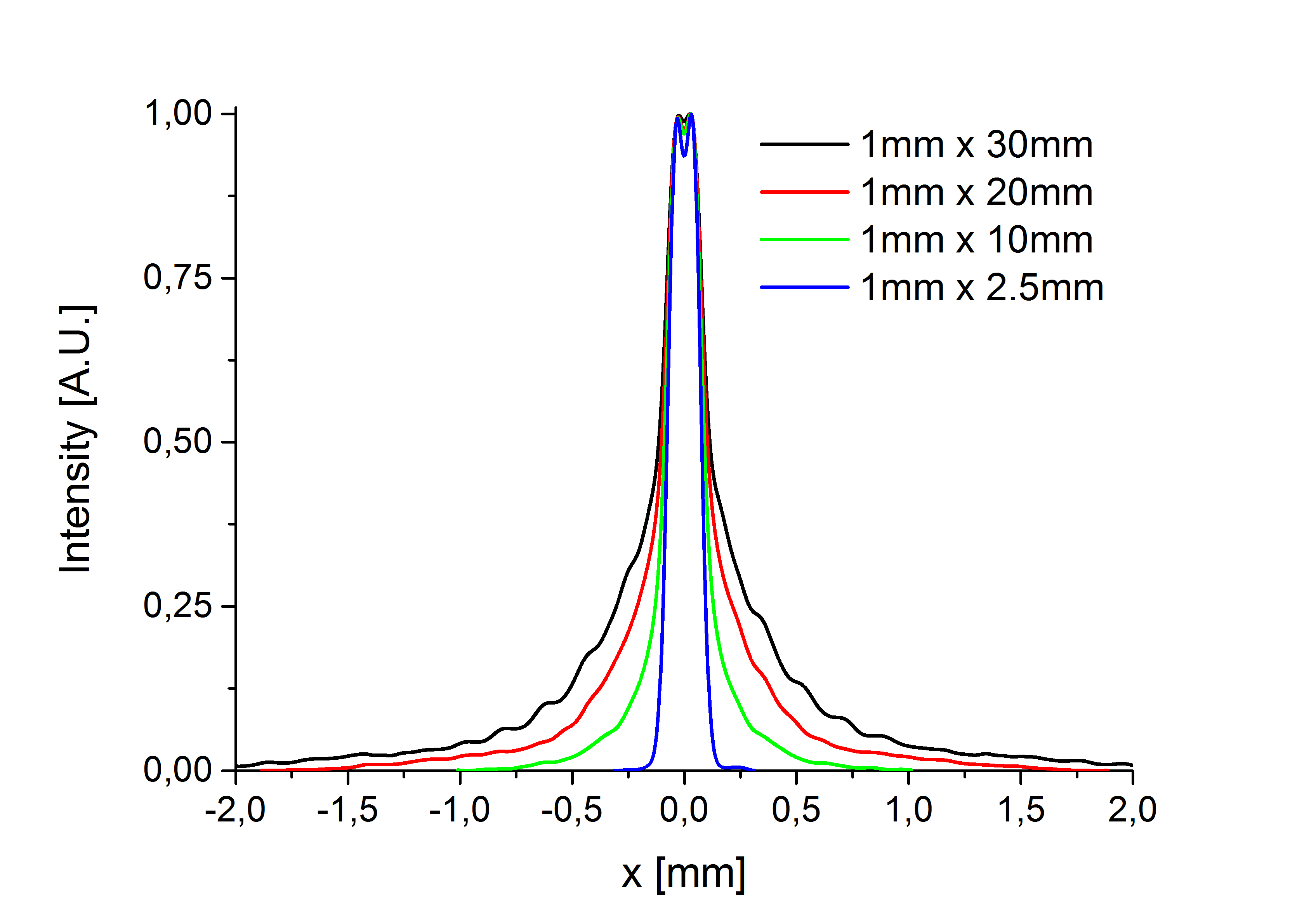}
\caption{Ideal wiggler. Horizontal cut at the median plane of the intensity profile at the source. The image is obtained simulating a perfect lens immediately behind a centered slit with varying horizontal size at 10 m from the middle of the wiggler. The slit considered here is $1$ mm wide in the vertical direction. The data correspond to a horizontal emittance $\epsilon_x = 1$ nm and a  horizontal beta function $\beta_x = 1$ m. In the vertical direction $\beta_y = \beta_x$, and $\epsilon_y = \epsilon_x/100$.  } \label{rmsideal}
\end{figure}

\begin{figure}[tb]
\includegraphics[width=0.5\textwidth]{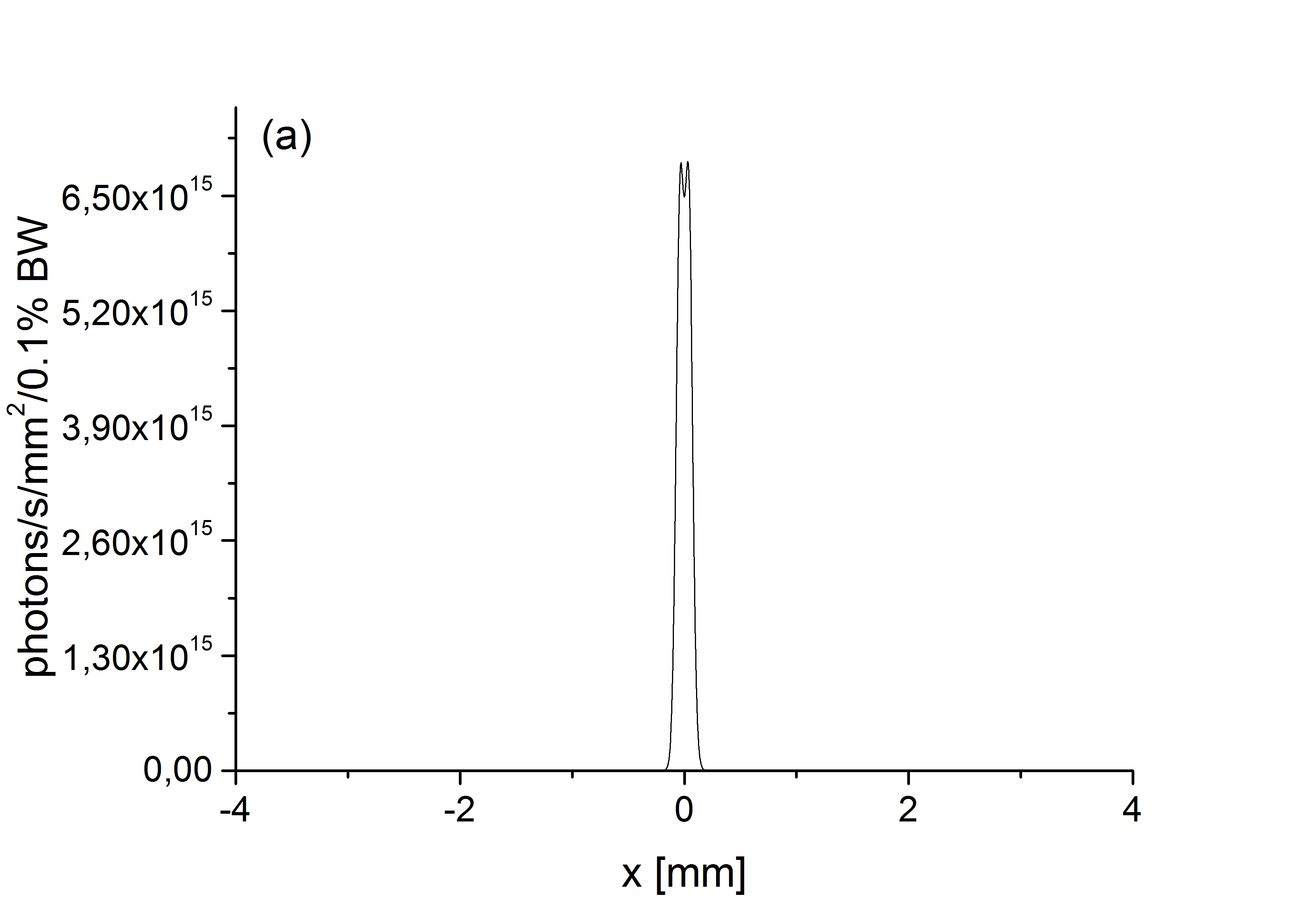}
\includegraphics[width=0.5\textwidth]{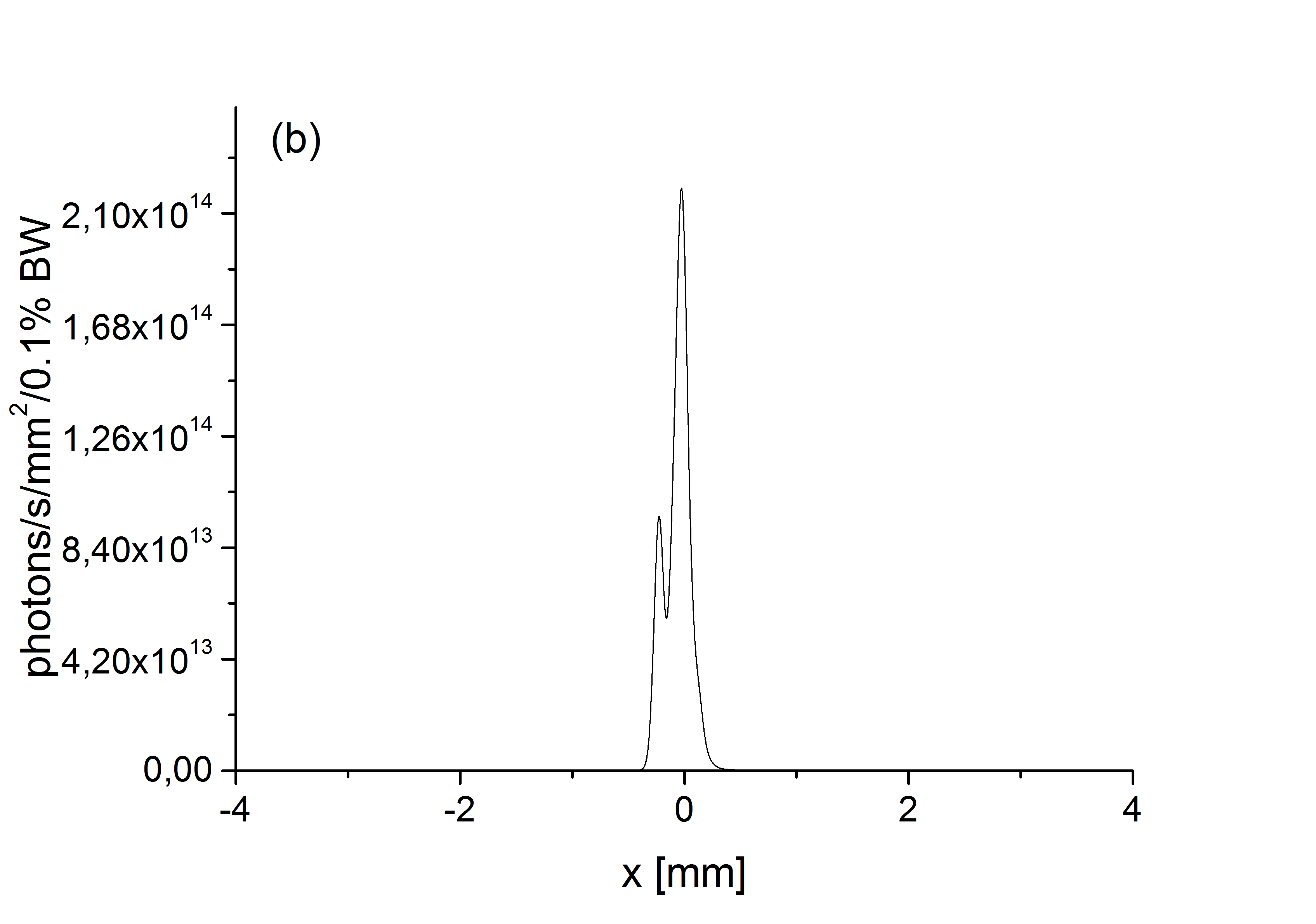}
\includegraphics[width=0.5\textwidth]{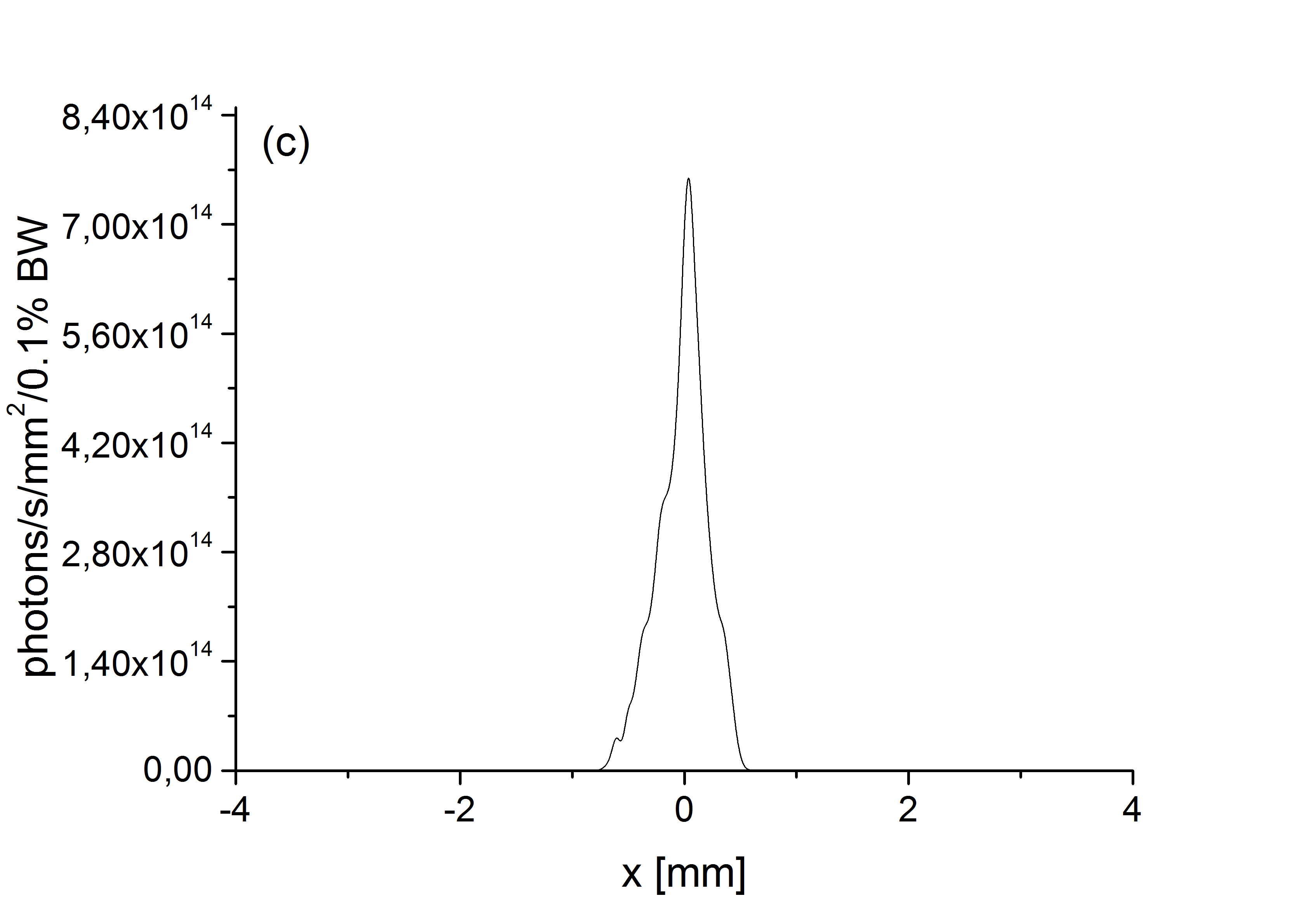}
\includegraphics[width=0.5\textwidth]{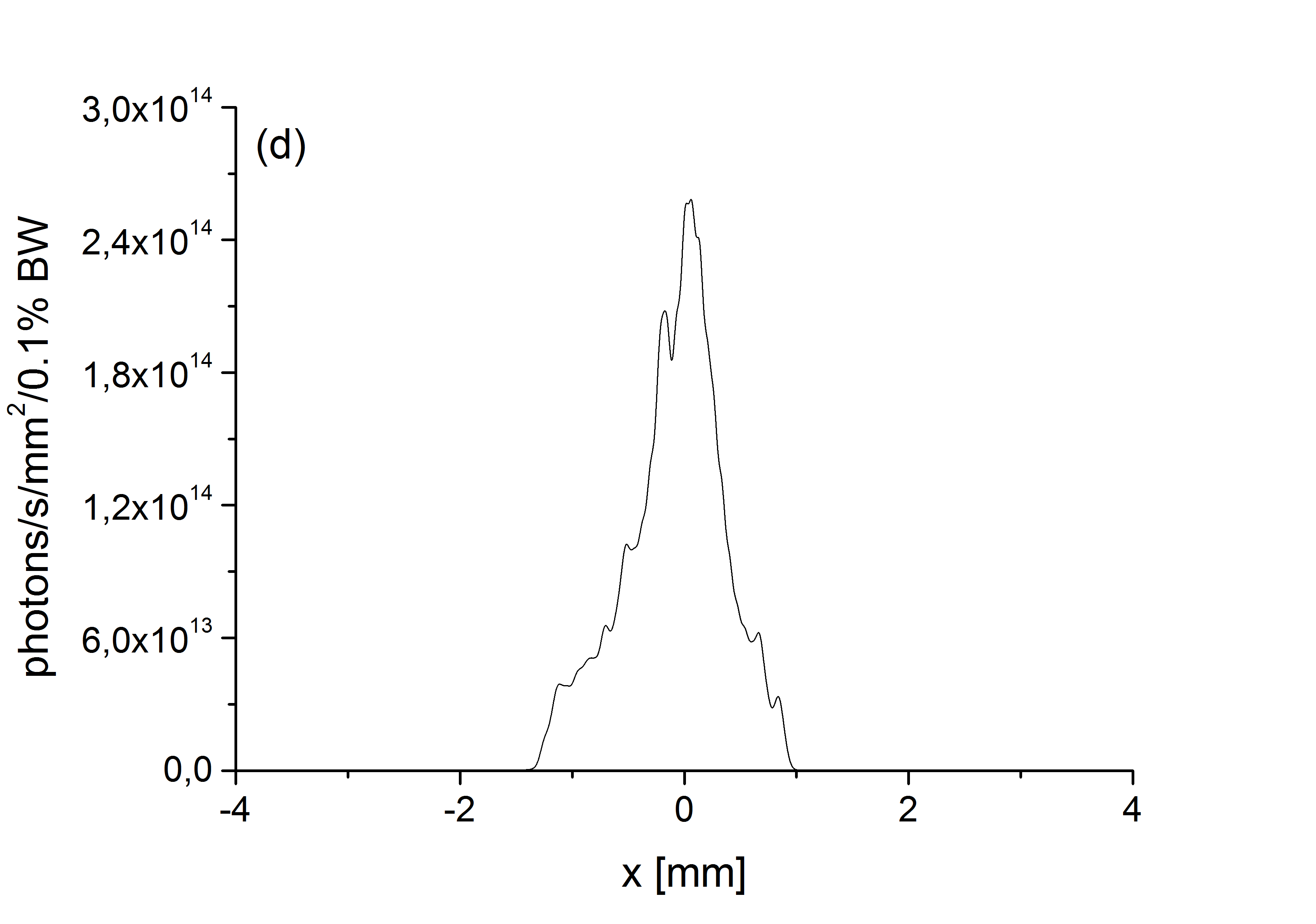}
\caption{Ideal wiggler. Horizontal cut at the median plane of the intensity profile at the source. The image is obtained simulating a perfect lens immediately behind a slit placed at $10$ m from the middle of the wiggler with varying off-axis transverse positions: (a) on axis,  (b) 1 mm off-axis, (c) 3 mm off-axis and (d) 6 mm off-axis. The slit considered here is $1$ mm by $1$ mm in size. The data correspond to a horizontal electron beam emittance $\epsilon_x = 1$ nm, and a horizontal betatron function $\beta_x = 1$ m. In the vertical direction $\beta_y = \beta_x$, and $\epsilon_y = \epsilon_x/100$. } \label{offset}
\end{figure}

Let us first consider the radiation emitted by an electron beam in an ideal wiggler, that is a  wiggler without field errors. Results from numerical simulations are presented in Fig. \ref{fieldfig}-\ref{offset}.

We first calculated the motion of an electron in the magnetic field of a wiggler, as shown in Fig. \ref{fieldfig}. The magnetic field variation along the $z$ axis of the wiggler can be approximately described by

\begin{eqnarray}
B_y(z) = B_w \sin\left(\frac{2\pi z}{\lambda_w}\right)~ .
\label{bfwig}
\end{eqnarray}
Two end-poles are included in the actual description of the field, in order to match the beam trajectory inside and outside the wiggler magnet. The motion of the reference electron is shown in Fig. \ref{trajfig}, while the electron beam and wiggler parameters are summarized in Table \ref{tab1}. The maximum transverse excursion is about $40 \mu$m. It is also instructive to calculate the maximum deflection angle for the reference electron. Since the $K$ value for the wiggler under study is $K= 12.3$, for $3$ GeV electrons we expect a maximum deflection angle $\theta_\mathrm{max} = K/\gamma \sim 2$ mrad.

The spectrum of an ideal wiggler is composed of the contribution of discrete harmonics. The on-axis frequency dependency of the wiggler photon flux density around resonance at the $101$st harmonic is shown in Fig. \ref{specid}. Intriguingly, this spectrum is visibly non-symmetric. We explain this behavior as an effect of the presence of the wiggler end-poles.

Next, computations of the spatial distribution are shown. In particular, Fig. \ref{fluxideal} - Fig. \ref{fluxemit_3D} illustrate the  flux distribution $10$ m away from the center of the wiggler for a filament electron beam (without emittance) and for a realistic electron beam (with emittance) at a fixed photon energy of $0.938$ keV corresponding, as discussed above, to the $101$st harmonic. Here we assume, for simplicity, that the electron beam waist is located in the middle of the wiggler. The resulting horizontal spot size is in good agreement with the above-made estimation of the maximum deflection angle $\theta_\mathrm{max} \sim 2$ mrad. The vertical rms spot size is in good agreement with the open angle estimation for a bending magnet, $\sigma_r' \sim (\lambda/2\pi R)^{1/3}$, where $R$ is the radius of curvature of the electron orbit in the bending magnet with magnetic strength $B= B_w$.

%
%
%

The central narrow bright peak and  substructures in Fig. \ref{fluxideal} and Fig. \ref{fluxideal_3D}  are a consequence of the periodicity of the field. The presence of electron beam emittance tends to broaden the peak and smooth the substructures. As discussed in section \ref{sec:wigwig}, for an electron beam with finite emittance the angular spectral flux is a convolution of the angular spectral flux produced by a filament beam with the angular distribution of the electron beam. Fig. \ref{fluxemit} and Fig. \ref{fluxemit_3D} present the result of such convolution. The main peak and some of the substructures are still  above the background level.

The intensity distribution at the 1:1 image plane provided by a perfect focusing lens for the case of a filament electron beam is presented in Fig. \ref{intevirt1} and Fig. \ref{intevirt1_3D}. We assume that the lens is placed $10$ m far away from the center of the wiggler, and that just before it a rectangular aperture of $2.5$ mm by $2.5$ mm limits the view. In the horizontal direction we find two separate sources for the case of wiggler radiation. This is ascribed to the horizontal shift of the electron trajectory, described above. The two sources correspond to the emission from positive and negative poles, separated by a distance $2a = 2 \lambda_w K/(2\pi\gamma) \sim 80 \mu$m. The presence of two source points  for wiggler radiation has been pointed out in \cite{WALK}.  Since the insertion device consists of $N$ periods in a length $L$, similar to the undulator case, the interference condition for an ideal wiggler can supply further information about the source size. The diffraction-limited source size depends on the wavelength and on the total length of the device, $L$. With this we get a source size of about $\sqrt{\lambda L/(2\pi)} \sim 26~ \mu$m. This result is in qualitative agreement with numerical results shown in Fig. \ref{slitvar}. Note that increasing the horizontal slit size has the effect of increasing the resolution of fine interference structures and source tails, as is clear by inspection of Fig. \ref{slitvar}.

The effect of the electron beam emittance on the intensity distribution at the 1:1 image plane is illustrated in Fig. \ref{intevirt_emit} and Fig. \ref{intevirt_emit_3D}, and can be compared directly with the result given earlier in Fig. \ref{intevirt1} and \ref{intevirt1_3D} for  an electron beam with zero emittance. We can see that the finite electron beam size has the predictable effect of spreading the width of the two sources and of smoothing out the fine interference structure present in the filament-beam case.


Numerical calculations also allow to verify that the `depth-of-field effects' does not exist. This is illustrated by comparing simulations for two different electron beam parameters at fixed electron beam size:

\begin{itemize}
\item A `realistic' case with horizontal emittance $\epsilon_x = 1$ nm and $\beta_x = 1$ m

\item An `extreme' case with tenfold increase in emittance and a tenfold decrease in beta function i.e. $\epsilon_x = 10$ nm, $\beta_x = 0.1$ m.
\end{itemize}

The results of computations are shown in Fig. \ref{horcutid}, and confirm our prediction that the image of the source is insensitive to the electron beam divergence. In both cases, with graphical accuracy, we find that the distribution at the image plane is the same, in contrast with the prediction in Eq. (\ref{Sigtotlit}) of an important source widening, for the second case, up to $0.6$ mm FWHM.

We now focus our attention on the apparent horizontal source size calculated as a function of the horizontal emission angle. These results are shown in Fig. \ref{rmsideal} where the intensity distribution at the 1:1 image plane has been calculated for different horizontal sizes of a centered slit. It can be seen that increasing the angular acceptance $\Theta$ up to a maximum $K/\gamma \sim 2$ mrad  practically has no effect on the FWHM,  due to the non-Gaussian form of the distribution. However, the horizontal source size variation calculated using the approximation in Eq. (\ref{Sigtotlit}) (and assuming a Gaussian form for the effective source profile) shows an increased FWHM of the source up to $2.1$ mm averaging over the range of horizontal emission angles\footnote{It should be noted that in \cite{WALK} the estimate in Eq. (\ref{Sigtotlit}) is interpreted as `mean square value' and it is mentioned about the non Gaussian form of the distribution. However, in practical work,  beamline scientists  usually convert the mean square value in Eq. (\ref{Sigtotlit}) to a FWHM  value assuming a Gaussian form for the effective source profile, see e.g. \cite{BERM}. }  $\pm 2$ mrad.

The off-axis radiation emission characteristics from wiggler sources have been described in \cite{WALK} with the help of the phase-space method.  A very important `blurring' of the effective horizontal source size when the wiggler radiation source is viewed off-axis, horizontally, is expected. Results of our simulations confirm this prediction. The effect of moving a slit defining a fixed angular acceptance off-axis is shown in Fig. \ref{offset}, where the intensity distribution in the 1:1 image plane has been calculated with fixed slit size ($1$ mm by $1$ mm), but with varying off-axis horizontal positions. The strong increase in the source size (up to $2$ mm FWHM at $6$ mm off-axis slit position)  can be clearly seen by inspection.

\subsection{Field error effects}

The quality of the magnetic field of a wiggler is often characterized by its deviation from the ideal case. A common way to characterize the magnitude of these deviations is simply to use rms peak field error as a figure of merit. The effect of field errors increases with the harmonic number. It is possible to reduce the field errors to the extent that operation in the ideal field approximation can be granted up to the $10$th-$15$th harmonic. Therefore, at the $101$st harmonic, the wiggler cannot be regarded as an ideal device anymore, and produces a `bumpy' but almost continuous spectrum.

\begin{figure}[tb]
\includegraphics[width=0.5\textwidth]{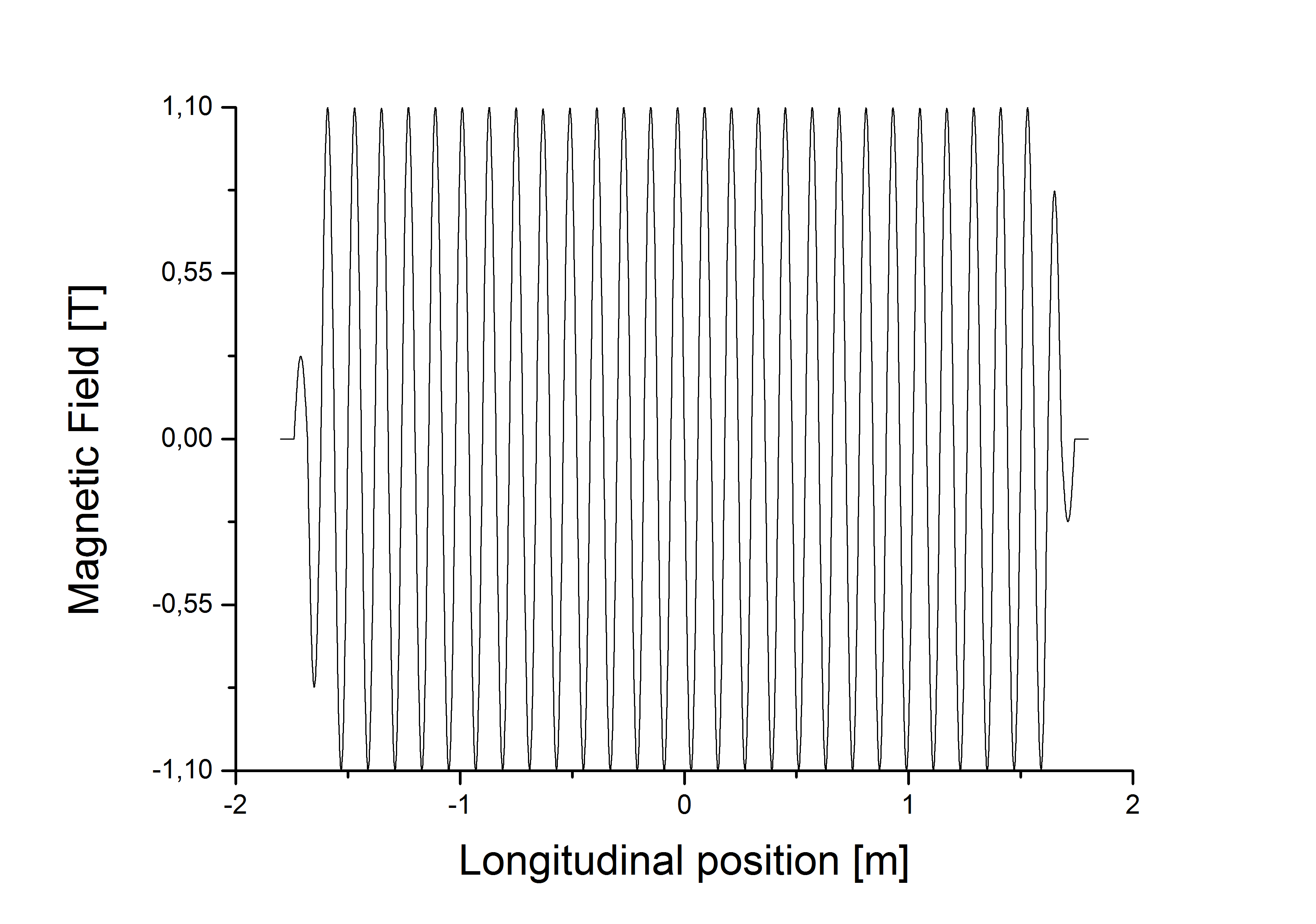}
\includegraphics[width=0.5\textwidth]{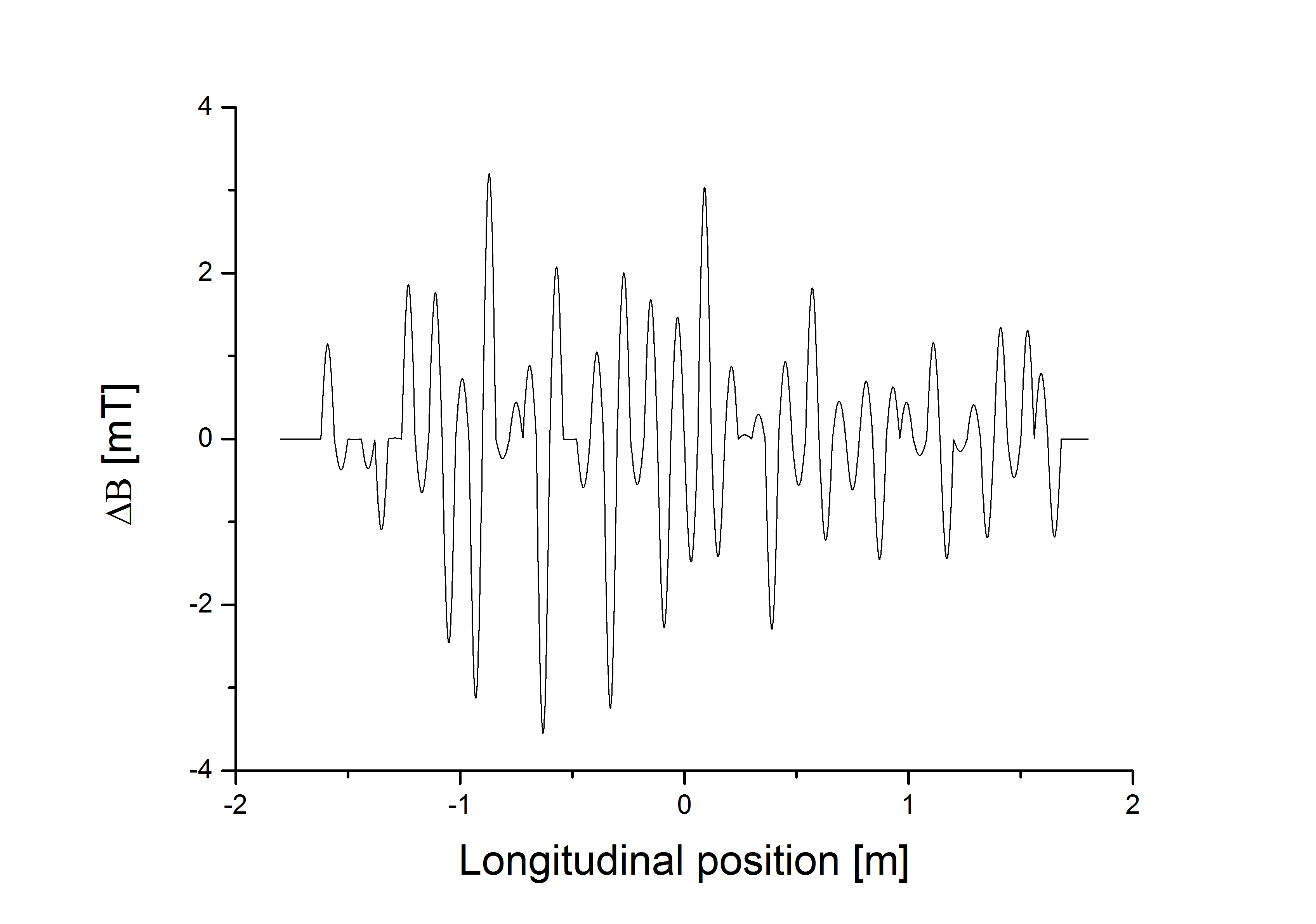}
\caption{Left: Magnetic field for a planar wiggler with random $0.1 \%$ normal distribution field errors included. Right: Plot of the difference $\Delta B(z) = B_\mathrm{non~ideal} - B_\mathrm{ideal}$.} \label{nonideal_field}
\end{figure}

\begin{figure}[tb]
\includegraphics[width=0.5\textwidth]{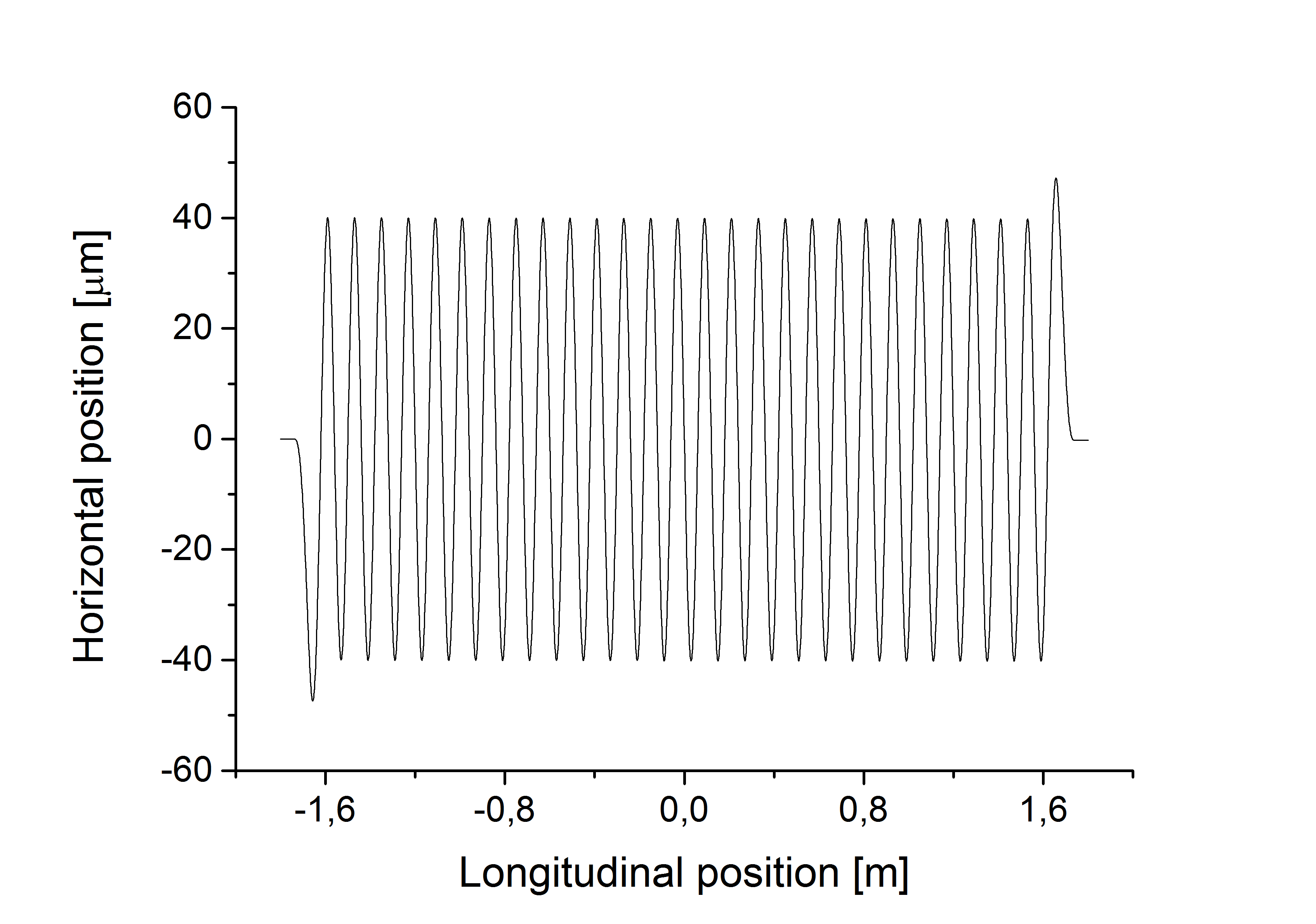}
\includegraphics[width=0.5\textwidth]{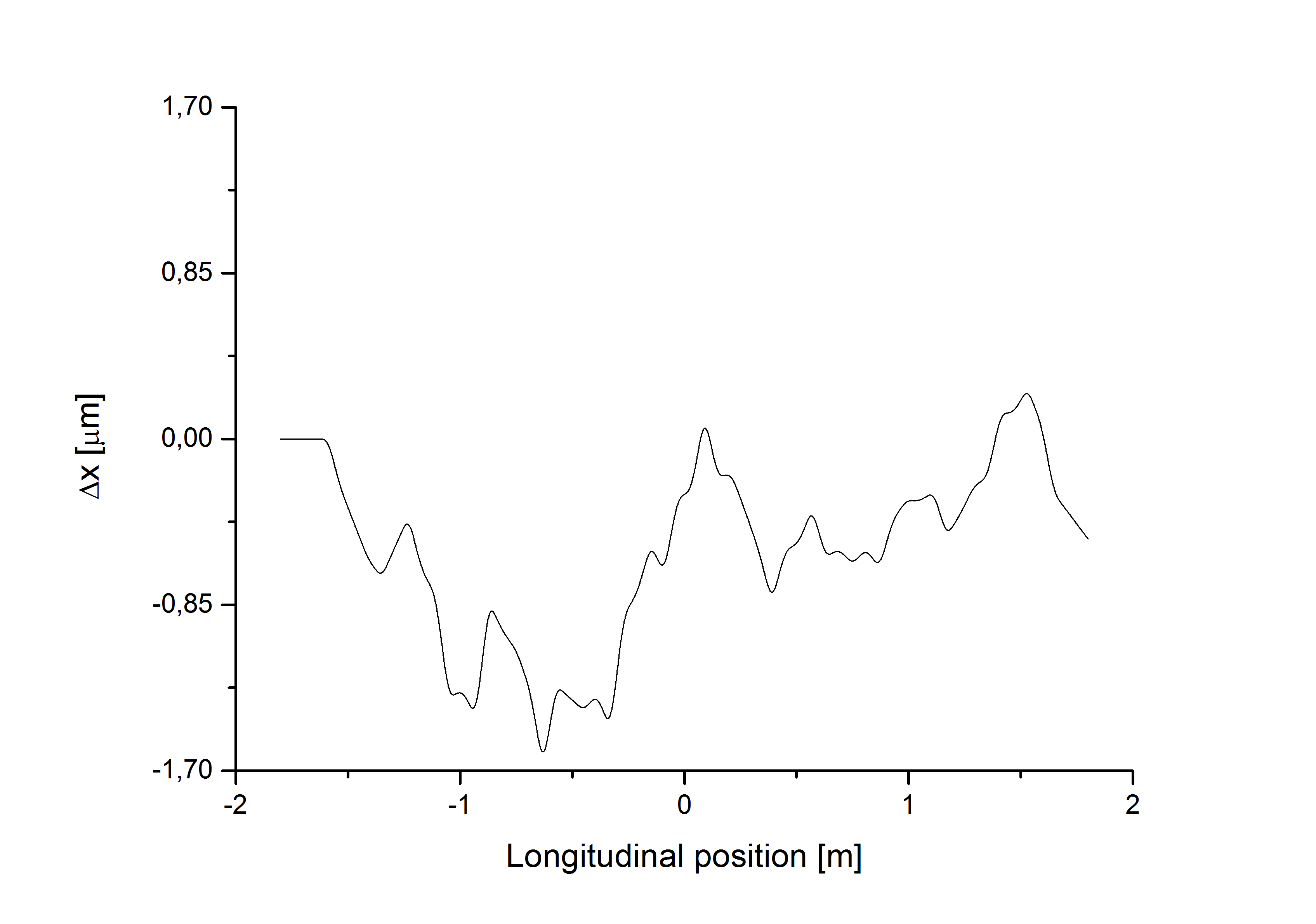}
\caption{Left: Trajectory of the $3$ GeV reference electron through a wiggler with field errors included as for Fig. \ref{nonideal_field}. Right: Plot of the difference $\Delta x(z) = x_\mathrm{non~ideal} - x_\mathrm{ideal}$.} \label{nonideal_traj}
\end{figure}

\begin{figure}[tb]
\includegraphics[width=1.0\textwidth]{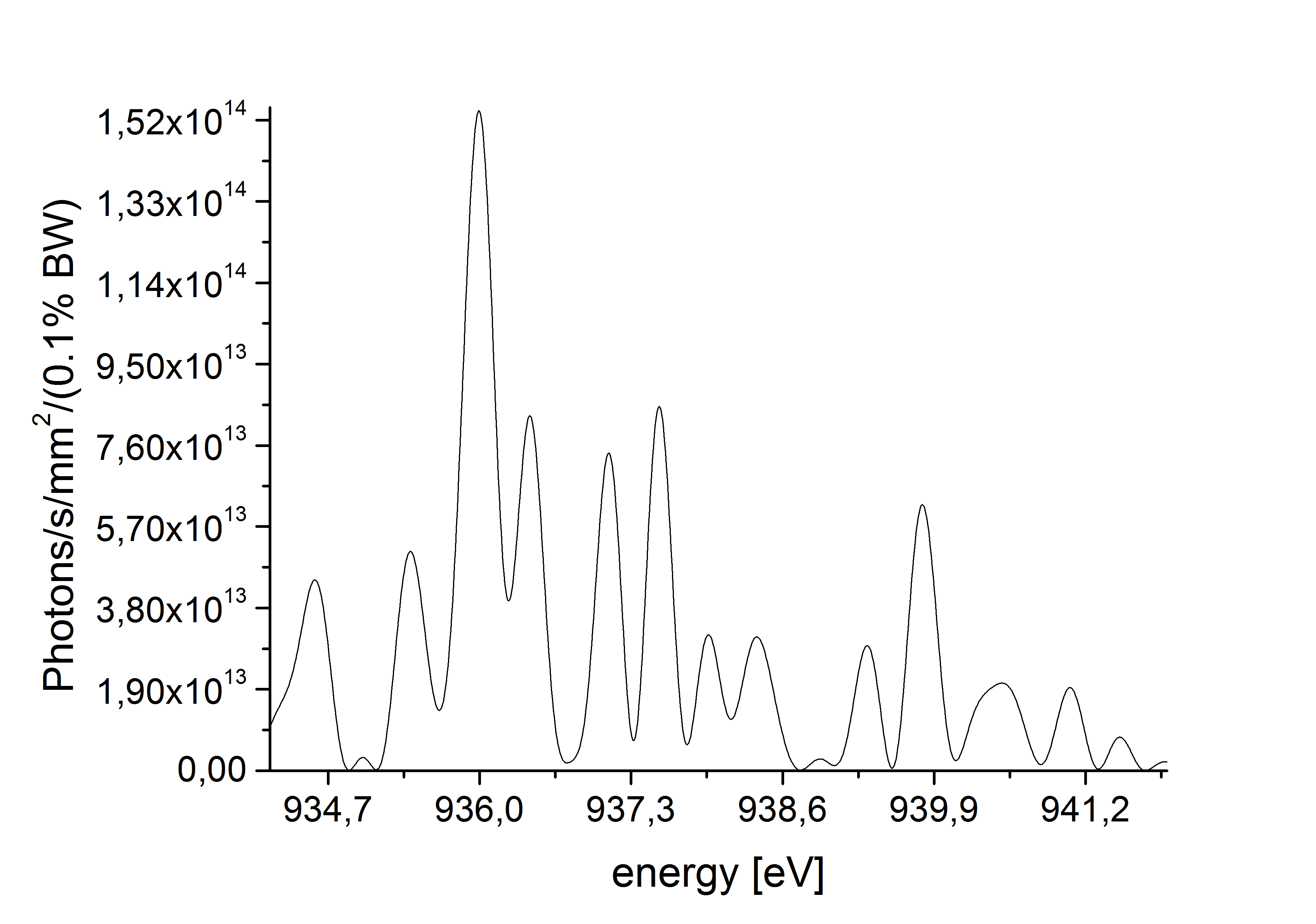}
\caption{Spectrum through a centered slit (with dimensions 0.1 mm by 0.1 mm) placed  at $10$ m from the middle of the wiggler with field errors included as for Fig. \ref{nonideal_field}. The flux density is shown as a function of the photon energy around the $101$st harmonic for an electron beam with zero emittance and energy spread.} \label{nonideal_spec}
\end{figure}

We begin our analysis of field error effects by  calculating the motion of a reference electron injected on-axis in the magnetic field of a non ideal wiggler such as that in Fig. \ref{nonideal_field}, where we simulate field errors of $0.1 \%$ rms. The motion of the electron is shown in Fig. \ref{nonideal_traj}.  The frequency dependence of the on-axis wiggler photon flux density around the photon energy of $101$st harmonic  is shown in Fig. \ref{nonideal_spec}. It can be seen that the interference structure is smoothed out, compared with the ideal case without field errors illustrated in Fig. \ref{specid}.

\begin{figure}[tb]
\includegraphics[width=1.0\textwidth]{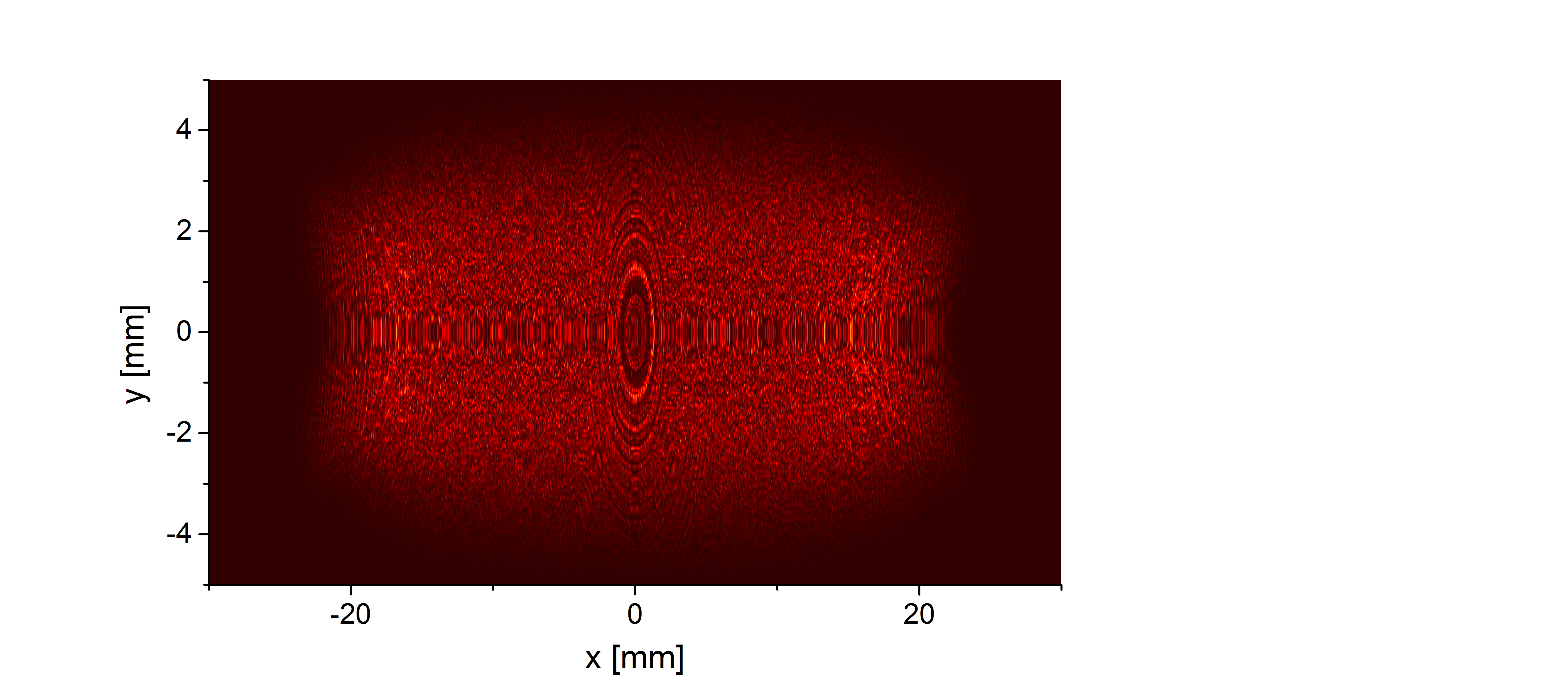}
\caption{Transverse flux distribution at 10 m from the middle of the wiggler with field errors included as for Fig. \ref{nonideal_field}. Emittance and energy spread are not included.} \label{nonideal_2dfar}
\end{figure}

It is also interesting to look at how the flux density varies with angles, both with and without wiggler field errors effects. Fig. \ref{nonideal_2dfar} shows the flux distribution $10$ m away from the center of the wiggler with imperfections for a filament electron beam, and can be directly compared with Fig. \ref{fluxideal} for the ideal case. It can be seen that the inclusion of amplitude field errors at a level of $0.1 \%$ rms has the effect of smoothing out the interference structure. The central peak and substructures are barely visible above the background level. This is compensated by an overall increase in the background level, which becomes larger  with the field errors.

\begin{figure}[tb]
\includegraphics[width=1.0\textwidth]{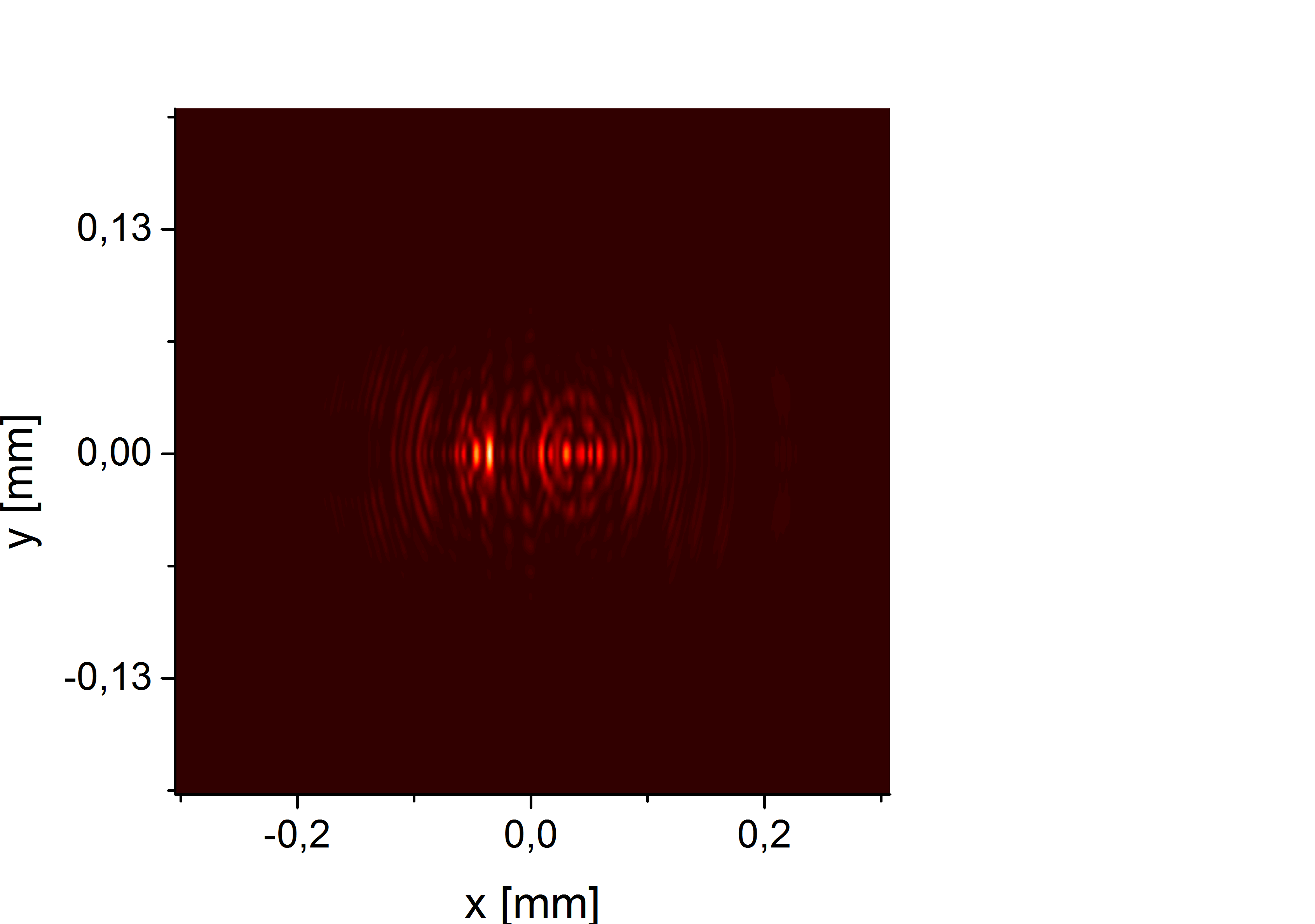}
\caption{Intensity distribution at the virtual source placed in the middle of
the wiggler. The image is obtained simulating a perfect lens immediately
behind centered rectangular aperture of 2.5 mm by 2.5 mm, placed at 10 m
from the middle of the wiggler. Field errors are included as for Fig. \ref{nonideal_field}. Emittance and energy spread are not included.} \label{nonideal_virt}
\end{figure}

\begin{figure}[tb]
\includegraphics[width=1.0\textwidth]{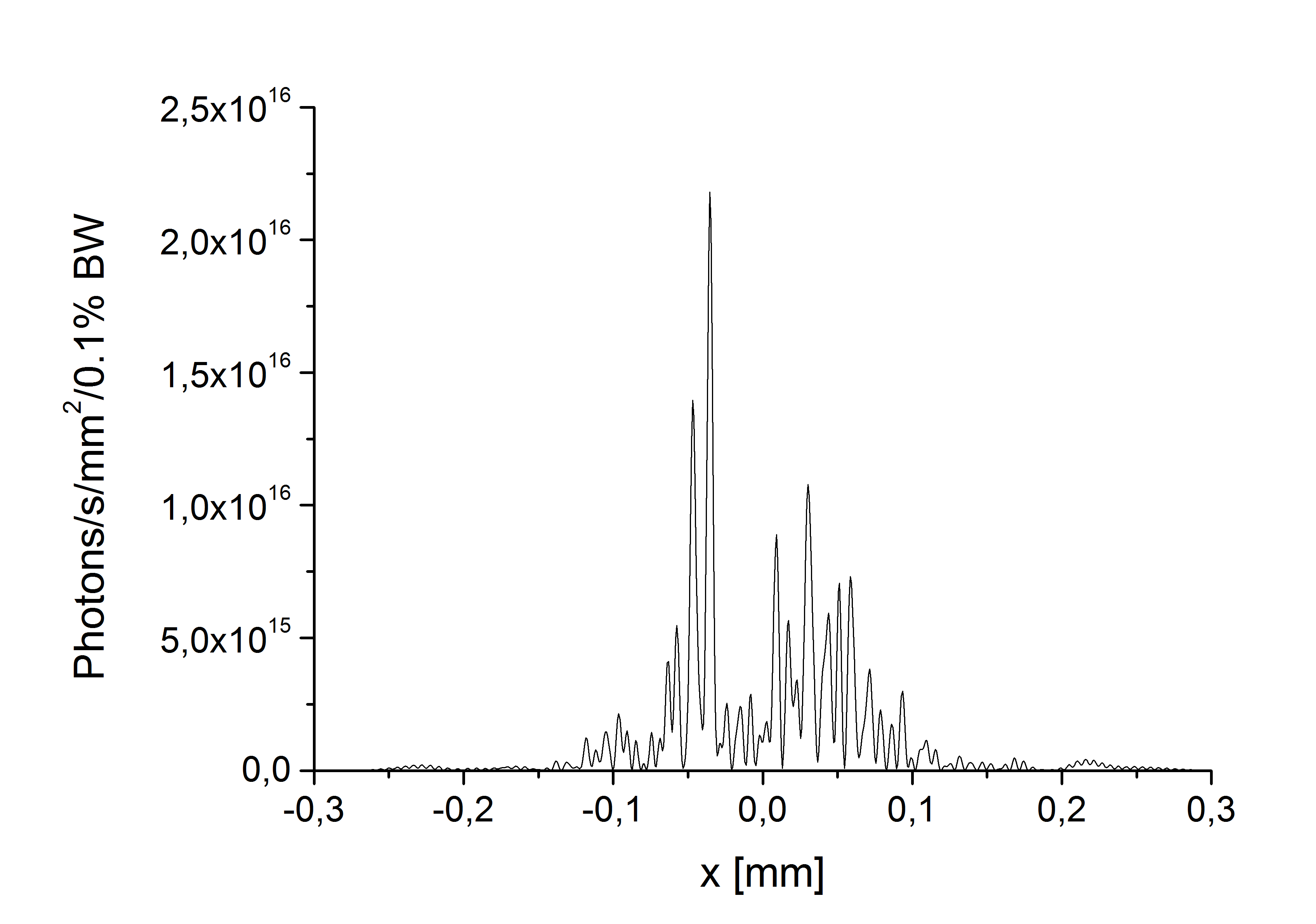}
\caption{Horizontal cut at the median plane ($y=0$) of Fig. \ref{nonideal_virt}.} \label{nonideal_graph6}
\end{figure}

The effect of field errors on the intensity distribution in the 1:1 image plane is presented in Fig. \ref{nonideal_virt} and Fig. \ref{nonideal_graph6} for the case of a filament beam.  These results can be directly compared with e.g. Fig. \ref{intevirt1} for an ideal wiggler. It can be seen that for a non ideal wiggler the interference structure is asymmetric and the two sources which correspond to the emission from positive and negative poles are significantly destroyed.

\begin{figure}[tb]
\includegraphics[width=1.0\textwidth]{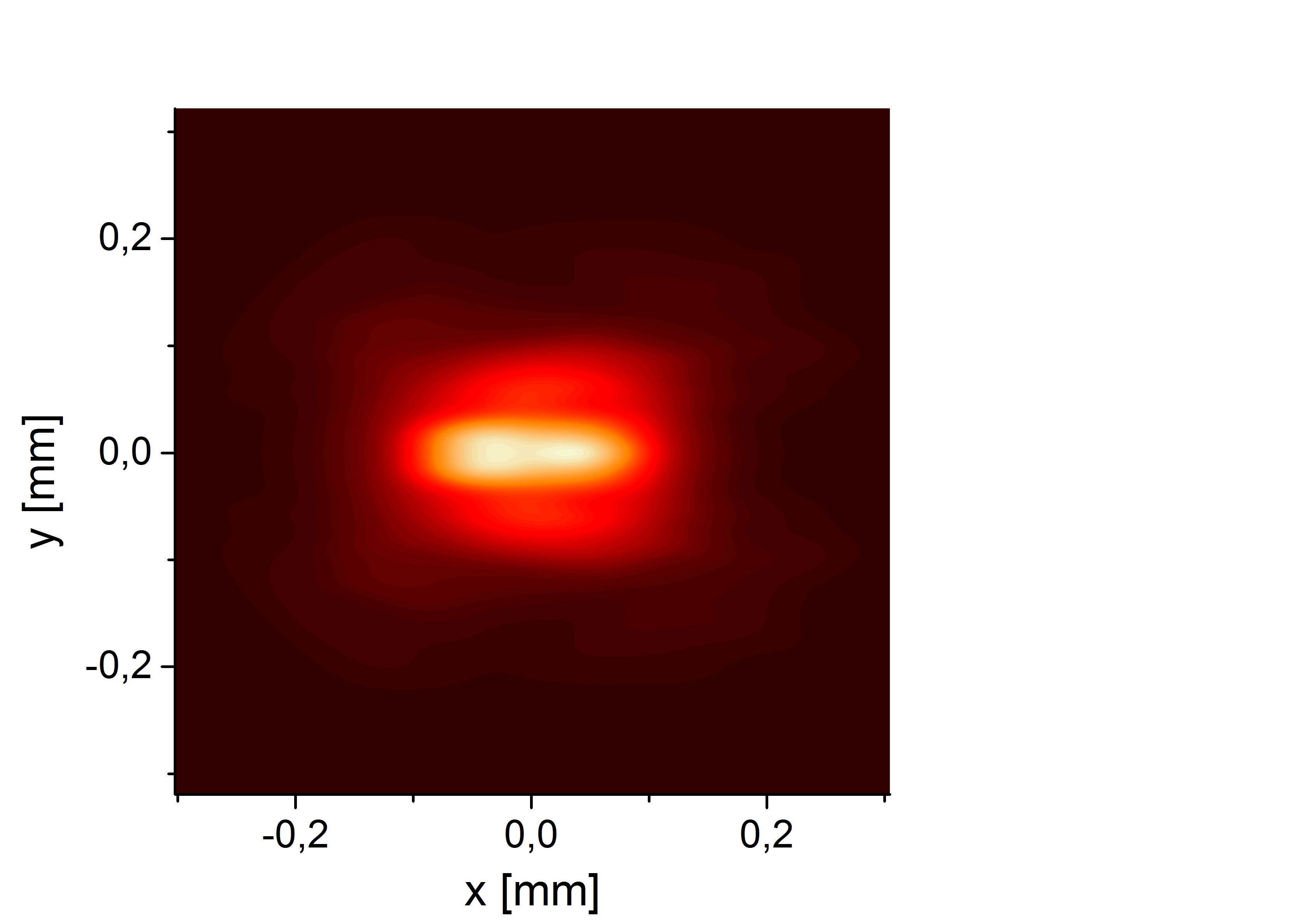}
\caption{Intensity distribution for the virtual source. The image is obtained simulating a perfect lens immediately behind a centered slit placed at $10$ m from the middle of the wiggler, with dimensions $2.5$ mm by $2.5$ mm. Field errors are included as for Fig. \ref{nonideal_field}.  The data correspond to a horizontal electron beam emittance  $\epsilon_x = 1$ nm, while $\beta_x = 1$ m. In the vertical direction $\beta_y = \beta_x$, and $\epsilon_y = \epsilon_x/100$.} \label{nonideal_2d}
\end{figure}

\begin{figure}[tb]
\includegraphics[width=1.0\textwidth]{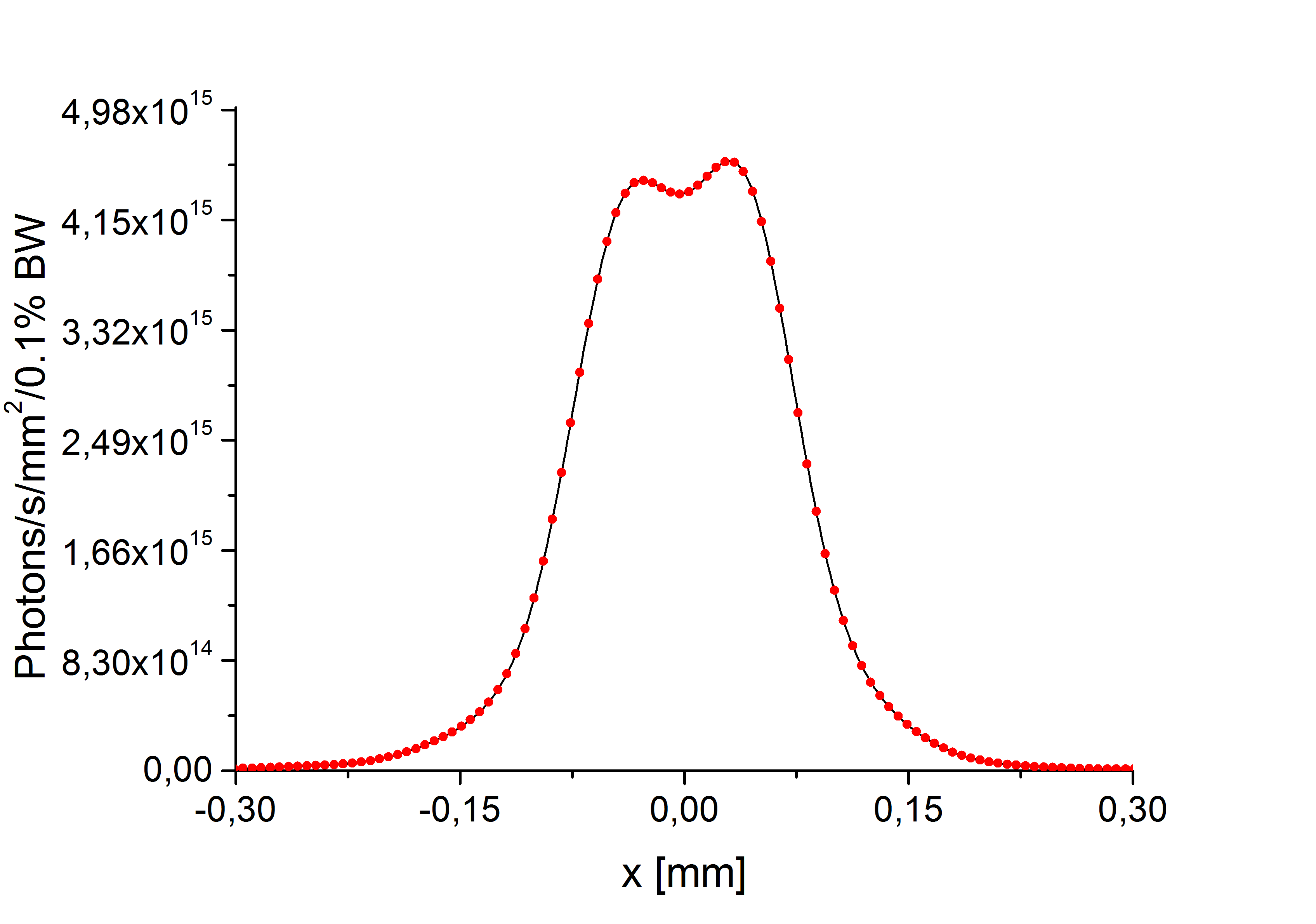}
\caption{Horizontal cut at the median plane of the intensity profile of the source of the wiggler with field errors included as for Fig. \ref{nonideal_field}. The image is obtained simulating a perfect lens immediately behind a centered slit placed at $10$ m from the middle of the wiggler, with dimensions $2.5$ mm by $2.5$ mm. The solid line corresponds to $\epsilon_x = 1$ nm, $\beta_x = 1$ m, and can directly obtain from Fig. \ref{nonideal_2d}, while the circles correspond to $\epsilon_x = 10$ nm, $\beta_x = 0.1$ m. In the vertical direction, for both cases, $\beta_y = \beta_x$, and $\epsilon_y = \epsilon_x/100$.} \label{nonideal_horcutid}
\end{figure}

Here we also provide an example the effects of a finite emittance on the calculated intensity distribution at the 1:1 image plane. Fig. \ref{nonideal_2d} and Fig. \ref{nonideal_horcutid} show the source shape obtained from simulations including the effect of field errors and electron beam emittance. This result can be compared directly with those in Fig. \ref{intevirt_emit}  and Fig. \ref{horcutid} for the ideal wiggler case. It can be seen that  the interference structure is smoothed out and the FWHM of the source shape in Fig. \ref{nonideal_horcutid} remains practically unchanged, compare to the case with zero field errors. Nevertheless, the source shape in Fig. \ref{nonideal_horcutid} is not symmetric due to wiggler imperfections.

Finally, we checked that `depth-of-field' effects are absent also in simulations with field errors. This can be  seen by inspection of  Fig. \ref{nonideal_horcutid}: with computational accuracy the image of the source is insensitive to the electron beam divergence as is expected to be. We conclude, more in general, that these effects are a misconception for arbitrary magnetic setups, in all those cases when the magnetic field only depends, with good approximation, on the longitudinal coordinate only (i.e. when focusing elements are absent).

\section{\label{sec:conc} Conclusions}

This paper discusses the theory of brightness for wiggler sources on the basis of classical relativistic electrodynamics and statistical optics. We consider the brightness defined with the help of a Wigner distribution. The theory of wiggler brightness is much more involved than that for undulators. In the far zone, and for a frequency close to the on-axis resonant frequency, undulator radiation from a single electron exhibits an angular divergence much smaller than $1/\gamma$  and a wavefront with a quadratic phase curvature. In contrast, wiggler radiation, due to a much larger  $K$ value, is emitted over the entire horizontal angle $K/\gamma \gg 1/\gamma$ and exhibits, even without accounting for field errors, a very complicated wavefront in the far zone. These differences explain why the brightness from wigglers was never described, up to now, in a satisfactory way within the Wigner distribution formalism. In particular, in usually accepted approximations, the description of the wiggler brightness turns out to be inconsistent even qualitatively. At variance, the analysis of undulator brightness given by Kim was based from the very beginning on the Wigner distribution formalism. As a result, the approximations he found for describing the brightness of undulators are parametrically consistent with all exact results.

A very useful `addition theorem' was originally pointed out by Kim for the undulator case. This theorem can be summarized by saying that  electron offset and deflection correspond to an offset in position and angle of the corresponding Wigner distribution, and that the overall Wigner distribution can be found by addition over single-electron contributions. We indicate an improvement of this theorem to the case of arbitrary magnetic setups without focusing elements (i.e. under the assumption that the magnetic field does not depend on the transverse position) \cite{BAZA,TANA} .

The results derived by applying the addition theorem for a wiggler source is far from trivial. As matter of fact, they are in contrast with literature. As a cardinal example, we consider the idea that the calculation of the wiggler brightness must take into account `depth-of-field' effects. We show that these effects do not exist. The core of the problem is the fact that according to electrodynamics the radiation from a single electron is emitted coherently at each point of the trajectory and that therefore the wiggler source is diffraction limited. At variance, in literature, an estimation of the source properties for
diffraction limited radiation from a wiggler is discussed based on geometrical optics, see e.g. \cite{WALK} for a review. In principle, this is no mistake. In fact, in the particular case of a wiggler, the departure of the far field wavefront from the spherical form can be considered as a severe aberration. Although the radiation from a single electron is coherent, diffraction effects are negligible compared to aberration effects, and the beam focusing can be calculated with the help of geometrical optics considerations.

Further on, we take into account the more general case when the electron beam has a finite phase space distribution. In this case, the Wigner distribution in the middle of the wiggler can be presented (according to the before-mentioned addition theorem) as a convolution product between the electron beam phase space and the Wigner distribution for a single electron in the middle of the wiggler. Since the convolution only involves the electron beam  phase space distribution in the middle of the wiggler, it follows that depth-of-field effects cannot exist. Moreover, also when the photon phase space distribution from a single electron is described by using ray tracing and  is subsequently convolved with the electron phase space distribution in the middle of wiggler depth-of-field effects are absent. The arguments usually  considered to explain their presence  rely on the following two points:

\begin{itemize}
\item{Radiation is incoherently summed up over whole poles in the case  of a single electron.}

\item{In the case of a finite electron phase-space distribution, according to the incoherent model for single electron, radiation from each pole must be convolved with the electron phase space distribution at the pole position along the wiggler.}
\end{itemize}

Using these two arguments, calculations actually predict the presence of depth-of-field effects as an increase in the apparent source size in the middle of wiggler. In this paper we argue that the wording `incoherently summed' is per-se misleading, and may only be used with some abuse of language. Moreover, we hold the second argument as incorrect, because it  actually assumes that poles radiate statistically independently from each other. At variance, we stress the existence of only one statistical ensemble, which is related with the electron beam distribution in phase space.

Based on the Wigner distribution method we have a well-defined procedure for computing the brightness from  a wiggler source. Thus, in principle, the problem of determination of the brightness of a given SR setup is solved. In particular, in the beam size-dominated regime, when the electron beam size is larger than the diffraction-limited source size, calculations become simple, and it is possible to calculate the brightness analytically (see e.g. Eq. (\ref{B3})). When the beam size is comparable with the diffraction-limited source size  the situation becomes more complicated, and must be solved numerically. The Wigner distribution is a function of four variables $(x,y,\theta_x,\theta_y)$ and at least five parameters $\sigma_{x,y}$, $\sigma_{x',y'}$ and $\omega$. The calculation of the brightness consists in finding the maximum of the Wigner distribution. Although conceptually straightforward, this can constitute a rather challenging computational problem. In \cite{TANA} numerical methods to compute the Wigner distribution of synchrotron radiation from wiggler sources are explored in more detail. All the methods presented in \cite{TANA} have been recently implemented in the SR calculation code SPECTRA \cite{SPECTRA}.

\clearpage



\clearpage

%


\end{document}